\documentclass[conference]{IEEEtran}
\usepackage{ifpdf}
\usepackage{tcolorbox}

\usepackage{tikz}

\usepackage{color,colortbl,array,xspace}
\usepackage{todonotes}

\definecolor{og}{rgb}{0.0, 0.5, 0.0}

\newcommand{\sys}{{\sc BlindMI}\xspace} 
\newcommand{\syssvm}{{\sc BlindMI-1class}\xspace} 
\newcommand{\sysdiff}{{\sc BlindMI-diff}\xspace}

\usepackage{cite}

\usepackage{graphics}

\usepackage[cmex10]{amsmath}
\usepackage{algorithm}
\usepackage{algorithmicx}
\usepackage{algpseudocode}

\usepackage{amsmath,amssymb,amscd}

\usepackage{mdwmath}
\usepackage{mdwtab}

\usepackage{eqparbox}
\newcolumntype{M}[1]{>{\centering\arraybackslash}m{#1}}

\usepackage[tight,footnotesize]{subfigure}

\newenvironment{icompact}{
  \begin{list}{$\bullet$}{
    \itemindent -.05em
    \parsep 0pt plus 1pt
    \partopsep 0pt plus 1pt
    \topsep 2pt plus 2pt minus 2pt
    \itemsep 0pt plus 1.3pt
    \parskip 0pt plus 2pt
    \leftmargin 0.13in}
       }
{\normalsize\end{list}}

\usepackage{fixltx2e}
\usepackage{stfloats}
\usepackage[hyphens]{url}
\usepackage{tabularx}
\usepackage{multirow} 
\usepackage{booktabs}  
\usepackage{threeparttable}
\usepackage[colorlinks,
linkcolor=red 
]{hyperref}

\usepackage{array}
\usepackage{mathtools}
\newcolumntype{R}[1]{>{\raggedleft\arraybackslash}p{#1}}
\usepackage{diagbox}
\usepackage{pifont}

\newcommand{\highlight}[1]{\textcolor{blue}{#1}}

\usepackage{spverbatim}
\usepackage{soul}

\definecolor{blue}{rgb}{0.0, 0.0, 0.0}
\definecolor{blue1}{rgb}{0,0,1}

\begin{document}




%
\title{Practical Blind Membership Inference Attack via Differential Comparisons}



%

\author{
Bo Hui$^{\dagger*}$,
 Yuchen Yang$^{\dagger*}$,
 Haolin Yuan$^{\dagger*}$, Philippe Burlina$^{\ddagger}$, Neil Zhenqiang Gong$^\mathsection$ and Yinzhi Cao$^\dagger$ \\
$^\dagger$The Johns Hopkins University $^\ddagger$The Johns Hopkins University Applied Physics Laboratory $^\mathsection$Duke University\\
}

\IEEEoverridecommandlockouts
\makeatletter\def\@IEEEpubidpullup{6.5\baselineskip}\makeatother
\IEEEpubid{\parbox{\columnwidth}{
    Network and Distributed Systems Security (NDSS) Symposium 2021\\
    21-24 February 2021\\
    ISBN 1-891562-66-5\\
    https://dx.doi.org/10.14722/ndss.2021.24293\\
    www.ndss-symposium.org
}
\hspace{\columnsep}\makebox[\columnwidth]{}}


%

\maketitle

\begingroup\renewcommand\thefootnote{$^*$}
\footnotetext{The first three authors have equal contributions to the paper. }
\endgroup

\begin{abstract}
Membership inference (MI) attacks affect user privacy by inferring whether given data samples have been used to train a target learning model, e.g., a deep neural network. There are two types of MI attacks in the literature, i.e., these with and without shadow models.  The success of the former heavily depends on the quality of the shadow model, i.e., the transferability between the shadow and the target;
the latter, given only blackbox probing access to the target model, cannot make an effective inference of unknowns, compared with MI attacks using shadow models, due to the insufficient number of qualified samples labeled with ground truth membership information.




In this paper, we propose an MI attack, called \sys, which probes the target model and extracts membership semantics via a novel approach, called differential comparison.  The high-level idea is that \sys first generates a dataset with nonmembers via transforming existing samples into new samples, and then differentially moves samples from a target dataset to the generated, non-member set in an iterative manner. If the differential move of a sample increases the set distance, \sys considers the sample as non-member and vice versa. 

 \sys was evaluated by comparing it with state-of-the-art MI attack algorithms.  Our evaluation shows that \sys improves F1-score by nearly 20\%  when compared to  state-of-the-art on some datasets, such as Purchase-50 and Birds-200, in the blind setting where the adversary does not know the target model's architecture and the target dataset's ground truth labels. We also show that \sys can defeat state-of-the-art defenses. 

\end{abstract}



%

\section{Introduction} \label{intro}

Machine learning (ML), especially Deep Learning (DL), has achieved, or even surpassed, human-level performance on many critical areas, such as medical diagnosis~\cite{burlina2017automated,burlina2011automatic}, image and speech recognition~\cite{resnet,dense,speech,imageRec}, self-driving cars~\cite{car}, and natural language translation~\cite{bart}.  Despite this success, one major issue of DL models like deep neural networks (DNNs) has been their vulnerability to a variety of attacks~\cite{modelInversion,tramer2016stealing,stealingPara,oh2019towards}. A type of privacy-related attack---i.e., the focus of the paper---is  the membership inference (MI) attack~\cite{earliest,globalloss,NDSS,BOS_attack}, whereby an adversary infers whether a specific sample belongs to the training set of a given learning model, defined as a membership.  For example, an adversary can infer whether a specific disease image from a given hospital was used to train an artificial intelligent (AI) diagnostic system, thus potentially violating patients' protected health information (PHI) and Health Insurance Portability and Accountability Act (HIPAA) provisions.  \highlight{For another example, the inference of location data used in an AI recommendation system may leak users' past physical location, violating their privacy.}

The high-level intuition behind membership inference attacks is that the output probability distributions of a DNN model from, say for example, a Softmax layer, may vary between members and a non-members.
%
  While intuitively simple, 
 %
 %
   one major {\it challenge} to this idea is that an adversary, when only given blackbox access (\highlight{i.e., only having access to the output probability distribution}), needs to collect enough samples with output probabilities and \highlight{labeled as either} members \highlight{or} non-members to classify a new data sample \highlight{with unknown membership}.  On one hand, many existing MI attacks---e.g., the DNN-based from Shokri et al.~\cite{earliest}, the loss function-based from Yeom et al.~\cite{globalloss}, and another DNN-based with feature selections from Salem et al.~\cite{NDSS}---all adopt an offline shadow model \highlight{trained from a surrogate dataset, that provides ground truth information on whether a given sample is a member}. 
   %
    However, such shadow models differ from the real target model and thus the output probability distributions, though being similar, as noted by \highlight{prior studies on model transferability~\cite{model_transfer_1,model_transfer_2,model_transfer_3}}, are still different.  Therefore if the shadow model is drastically different from the target, the attack performance will degrade significantly as shown by Salem et al.~\cite{NDSS} and as also confirmed by our own experiments. 
    
   
 On the other hand, researchers have also proposed MI attacks without shadow models. 
 %
  For example, the unsupervised and adversary binary attacks of Salem et al.~\cite{NDSS} 
  consider a sample as a member if
  the probability of the predicted class is larger than a threshold learned from one thousand randomly generated samples. 
  Another approach, the label-only attack of Yeom et al.~\cite{globalloss}, infers membership by comparing the ground truth against the predicted label. However, both existing shadow-model-free attacks rely binary comparisons, e.g., comparing the predicted probability or label with a pre-determined threshold or the ground-truth label.  Such a ``one-size-fit-all'' inference cannot model the complex decision boundary between members and nonmembers in the hyper-dimensional space induced by an inference neural network in shadow-model-dependent attacks. 
 %
 %
   The root reason in lacking such modeling ability 
    goes back to the aforementioned challenge: a powerful MI attack needs enough labeled output probability distributions of members and non-members to learn the decision boundary, but the ground truth information of members and nonmembers for the target model is unavailable given only blackbox access.

In this paper, we propose a novel MI attack, called \sys, which probes the target model and then infers membership directly from the probing results instead of shadow models. 
%
 \sys exploits two insights:
 The first  is that although an adversary does not have both member and nonmember labels of the target model, the adversary can easily obtain one-class labels, i.e., nonmember labeled samples, by producing newly-constructed samples that can be considered as non-members with high probability given the very large input space of possibilities. Such one-class semantics can be learned by existing ML classifiers, like a one-class SVM, thus leading to an MI attack defined as \syssvm.  \highlight{This \syssvm serves as a baseline approach if we only exploits the first insight of \sys.}

The second insight is that the removal of a non-member from a dataset containing both members and non-members, will move the entire set away from non-members in the hyper-dimensional space, and conversely, the addition moves it towards it. Therefore, assume that we have two datasets: one is closer to nonmembers and the other to members.  If we move one sample from the latter to the former and the distance between two sets decreases, the moved sample can be considered a member; otherwise, if it increases, the sample can be considered a non-member.  This approach is called {\it differential comparison} in this paper as it compares the differential distance between two sets. One advantage of this approach is that it only needs two small-size sets as opposed to a considerable amount of data for a one-class classifier, while achieving a comparatively higher inference performance. 

Specifically, we design an attack, called \sysdiff, which performs differential comparison to infer membership. Following upon the first insight, \sysdiff obtains a dataset with nonmembers.  Then, \sysdiff differentially compares the dataset with a given set of data samples, called a target dataset, 
%
 following the second insight to remove all the nonmembers from the target.  The entire differential comparison procedure is iterated until convergence, i.e., the move of any samples between two sets only decreases the distance.  Then, the remaining samples in the target dataset are considered as members.

We implement a prototype of \sys\footnote{The default version of \sys and \sysdiff, without specification, is \sysdiff with generated non-members, called \sysdiff-w/.} including \sysdiff and \syssvm. Our evaluation shows that \sys outperforms state-of-the-art membership inference attacks in terms of F1-score in different settings, e.g., even when the adversary knows the target model's exact architecture and hyper-parameters. Furthermore, we evaluate \sys and other attacks under realistic assumptions following Bargav et al.~\cite{realisticMI} to adjust the nonmember-to-member ratio in the target dataset and show that even if the ratio is as high as 39, \sys still has an over 50\% F1-score as opposed to below 30\% of the state-of-the-art MI attacks. 
%
%
 We also test \sys against existing defenses, including Adversarial Regularization~\cite{minmax}, MemGuard~\cite{memguard}, Mixup $+$ MMD~\cite{mixup} and differential privacy~\cite{DP}, and show that \sys can break these defenses by achieving reasonable F1-score with different privacy-utility budgets.

\section{Overview}\label{Overview}

In this section, we first present our threat model and then describe overarching assumptions and principles  used  throughout the paper.  

\subsection{Threat Model}  \label{subsec:threatmodel}

\begin{table}[!t]
   \renewcommand{\arraystretch}{1.2}
   \setlength{\tabcolsep}{3pt}
   \footnotesize
    \caption{Different Threat Models and Their Assumptions.} \vspace{-0.05in}
    \label{tab:adversarysetting}
	\centering
	\begin{tabular}{lccc}
	    \toprule 
	    & \multirow{2}{*}{\bf output distr.} & {\bf model arch.\&}  & \multirow{2}{*}{\bf targets' true labels} \\
	    & & {\bf hyper-parameter}& \\
	    	    \midrule
	    Blind (default) & \ding{51} & \ding{55} & \ding{55}\\
	    Blackbox & \ding{51} & \ding{55} & \ding{51}\\
	    Graybox & \ding{51} & \ding{51} & \ding{51}\\
	    Graybox-Blind & \ding{51} & \ding{51} & \ding{55}\\
	    \bottomrule
	 \end{tabular}
\end{table}

Our threat model 
 assumes an adversary trying to infer whether each sample in a given input dataset, called the target dataset, belongs to---i.e., is a member of---the training set of a deep learning (DL) model, called the target model.
  The adversary can probe the target DL model with samples to obtain the probability distribution of output classes.  There are four different variations of the threat model based on the adversary's capability as described below and shown in Table~\ref{tab:adversarysetting}.

\begin{icompact}
\item Blackbox-Blind, or called Blind for short.  \hspace{0.05in} The blind setting only grants an adversary blackbox access to the target model without details of its architecture, network weights, or hyper-parameters.  Further, the adversary does not have ground truth class  labels of the target dataset, which usually takes a considerable amount of manual effort \highlight{sometimes even from specialized experts, e.g.,  a highly trained ophthalmologist and retinal specialist in labeling the existence of certain diseases for the EyePACS dataset.} 


\item Blackbox. \hspace{0.05in} The blackbox setting is similar to the blind, but assumes that the adversary has the ground-truth information of all the samples in the target dataset via, e.g., manual labelling.  Note that some existing attacks, e.g., Yeom et al.~\cite{globalloss}, only work if such ground-truth information is available.

\item Graybox. \hspace{0.05in} The graybox setting gives full knowledge to the adversary in terms of the model details. Specifically, except for the training data, the adversary knows almost everything about the model, such as the architecture (e.g., VGG, ResNet, and DenseNet) and the hyper-parameters used for training (e.g., learning rate and maximum number of epochs). Note that the adversary cannot know the training data (called a whitebox), because MI attacks are unnecessary in such a setting.  



\item Graybox-Blind. \hspace{0.05in} The graybox-blind setting is similar to the graybox one, but also assumes that the adversary does not have ground-truth information of the target dataset. 

\end{icompact}

Note that our {\it default} threat model setting is blind unless otherwise noted, because the blind setting is the most strict and practical for membership inference attack.  We also adopted other settings in comparison with prior works, e.g., blackbox with Yeom et al.~\cite{globalloss} and graybox(-blind) with Shokri et al.~\cite{earliest} and Salem et al~\cite{NDSS}.

\subsection{Problem Formulation and Notations}  \label{formulation}

The attack problem considered in this paper is as follows: given a target model, i.e., $F_m$ (see recapitulation of notations in Table~\ref{tab:notations}), with $m$ classes and a target dataset $S_{target}$, \sys is tasked with inferring whether each sample in $S_{target}$ belongs to the training set of $F_m$.  The adversary will feed the set of samples $S_{target}$ into $F_m$, obtain the set of output probability distributions, i.e., $S^{prob}_{target}$, and 
\textcolor{blue}{then applies a projection function $G_{projection,k}$ that converts all data samples from $m$ dimensions to $k$ dimensions for inference.  The converted samples form into a new set called $S^{prob,k}_{target}$}.  One example $G_{projection,k}$ is the selection of top three probabilities in $y\in S^{prob}_{target}$ plus the one corresponding to the ground truth class.  

Another important dataset is a generated dataset, $S_{nonmem}$, which is a reference dataset to determine whether samples in $S_{target}$ are members. 
 Elements in  $S_{nonmem}$ are all---or mostly---non-members of the target model training dataset. 
  Similarly, the adversary will also obtain $S^{prob}_{nonmem}$ and $S^{prob,k}_{nonmem}$ for the comparison with $S_{target}$.


\begin{table}[!t]
    \renewcommand{\arraystretch}{1.2}
    \caption{Notations of symbols in the paper} \vspace{-0.05in}
    \label{tab:notations}
	\centering
	\begin{tabular}{r|llc}
	    \toprule
        \multicolumn{2}{c}{\textbf{Notation}} & \textbf{Description} \cr
        \midrule
        \multicolumn{2}{c}{$S_{target}$} &  The target dataset of membership inference attack  \cr
          \multicolumn{2}{c}{$F_m$} & The target DL model with $m$ prediction classes    \cr
        \multicolumn{2}{c}{$S^{prob}_{target}$} &  $S^{prob}_{target}=\{y=\highlight{F_m}(x)|x\in S_{target}\}$  \cr
        \multicolumn{2}{c}{$G_{projection,k}$} & $G_{projection,k}: \mathbb{R}^m\rightarrow \mathbb{R}^k$      \cr
        \multicolumn{2}{c}{$S^{prob,k}_{target}$} &  $S^{prob,k}_{target}=\{y^\prime=G_{projection,k}(y)|y\in S^{prob}_{target}\}$  \cr

                \multicolumn{2}{c}{$S_{nonmem}$} &  A generated dataset with non-members of $F_m$  \cr
        \multicolumn{2}{c}{$S^{prob}_{nonmem}$} &  $S^{prob}_{nonmem}=\{y=\highlight{F_m}(x)|x\in S_{nonmem}\}$  \cr
        \multicolumn{2}{c}{$S^{prob,k}_{nonmem}$} &  $S^{prob,k}_{nonmem}=\{y^\prime=G_{projection,k}(y)|y\in S^{prob}_{nonmem}\}$  \cr
        
%
%
        \bottomrule
	\end{tabular}
\vspace{-0.05in}
\end{table}

\subsection{Differential Comparison Intuition} \label{intuition}

\begin{figure}[!t]
\centering
\includegraphics[width=0.5\textwidth]{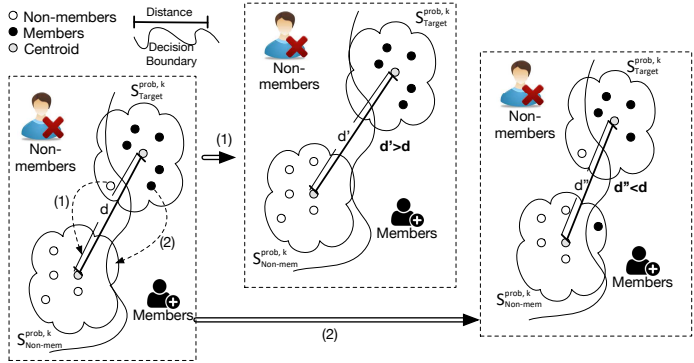} \vspace{-0.25in}
\caption{A High-level \highlight{Idea of the Key, Atomic Step in}  Differential Comparisons. \highlight{\sys measures the distance $d$ between $S^{prob,k}_{target}$ and $S^{prob,k}_{nonmem}$ and moves a sample from $S^{prob,k}_{target}$ to $S^{prob,k}_{nonmem}$. Then, \sys recalculates the distance $d'$ and compares $d'$ with $d$. If $d'$ is larger than $d$, \sys considers the moved sample as a nonmember; otherwise, \sys considers it as a member. This is an iterative process until convergence.}}
\vspace{-0.05in}
\label{fig:intuition}
\end{figure}

We now introduce the high-level idea, i.e., differential comparison, in Figure~\ref{fig:intuition}. We depict two datasets, $S^{prob,k}_{nonmem}$ and $S^{prob,k}_{target}$, in the output probability distribution sub-space. The curve dividing the space is the boundary between member (right) and non-member (left).  $S^{prob,k}_{nonmem}$ is located more towards the left because it consists exclusively of samples with high probability of being non-members, while $S^{prob,k}_{target}$ more or less in the middle between members and non-members. 

\highlight{Intuitively,} the idea of differential comparison is to move one sample from $S^{prob,k}_{target}$ to $S^{prob,k}_{nonmem}$.  If the moved sample is a non-member like Case (1) in Figure~\ref{fig:intuition}, $S^{prob,k}_{nonmem}$ moves further towards the left and $S^{prob,k}_{target}$ to the right.  Therefore, the distance between $S^{prob,k}_{target}$ and $S^{prob,k}_{nonmem}$ increases from the original $d$ to $d^\prime$. If the moved sample is a member like Case (2), the distance will decrease from $d$ to $d^{\prime\prime}$since both sets are now comprised of a mixture of samples. Such a change in $d$ can then be used to infer whether the moved sample is a member.

 \highlight{While intuitively simple, the distance between $S^{prob,k}_{nonmem}$ and $S^{prob,k}_{target}$ changes over time after each move.  That is, what we described in the previous paragraph is a key, atomic step of differential comparison. In practice, 
%
this atomic step is repeated until no more samples can be moved: The series of moves with a fixed $d$ is defined as one iteration. Then, differential comparison will update $d$ based on new $S^{prob,k}_{target}$ and $S^{prob,k}_{nonmem}$ for another iteration until the distance $d$ does not change across iterations, called convergence.} 

\highlight{There are two things worth noting here.  First, }differential comparison moves one sample instead of removing it so as to maximize the distance change. Removal of a sample only changes the position of $S^{prob,k}_{target}$ with regards to the decision boundary in the hyper-dimensional space (like Figure~\ref{fig:intuition}); as a comparison, moving the sample 
 changes the positions of both $S^{prob,k}_{target}$ and $S^{prob,k}_{nonmem}$, thus 
 improving the algorithm's sensitivity. 
  \highlight{Second, even after convergence, there may still exist some nonmembers left in $S^{prob,k}_{target}$, i.e., the moving of these samples does not increase the distance between  $S^{prob,k}_{target}$ and $S^{prob,k}_{nonmem}$.  This is a lower probability situation, as shown in our evaluation of differential comparison's performance, and that this is as to be expected due to an inherent ambiguity between members and non-members.}

\section{Design} \label{Proposed New Attack}

In this section, we describe a detailed design of \sys.

\subsection{Overall Attack Procedure} \label{overall}


\begin{figure*}[!t]
\centering
\includegraphics[width=\textwidth]{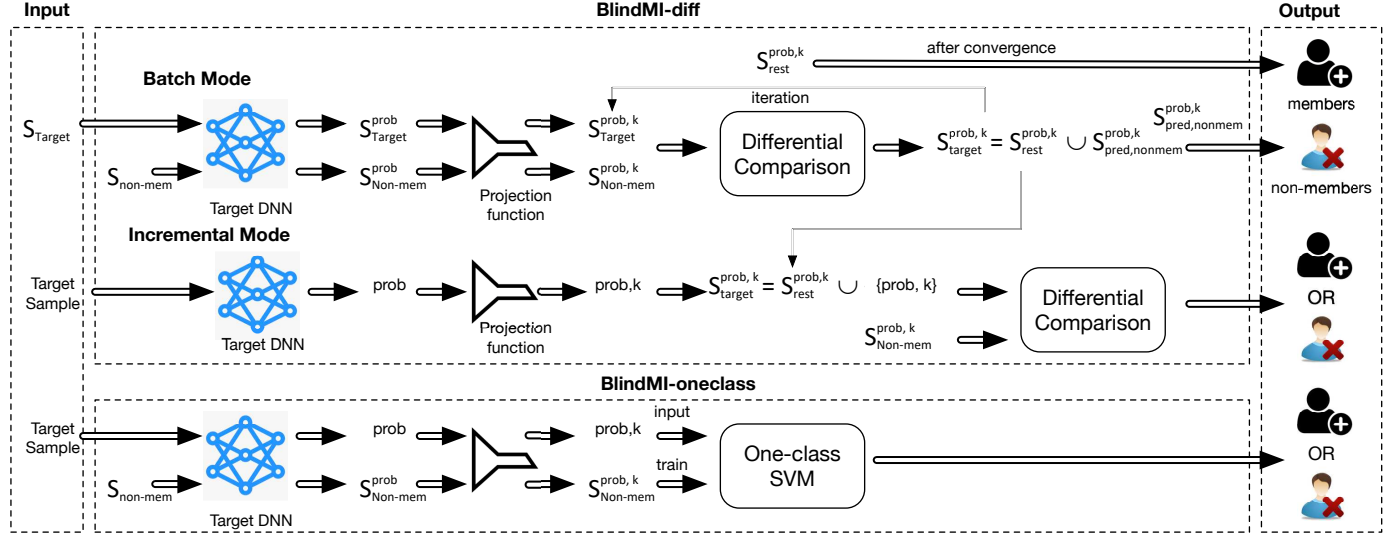}\vspace{-0.15in}
\caption{Overall Attack Procedure of \sys} \vspace{-0.1in}
\label{fig:arch}
\end{figure*}

\begin{table}[!t]
 \renewcommand{\arraystretch}{1.2}
 \setlength{\tabcolsep}{2pt}
 \scriptsize
    \caption{Different Variations of \sys} \vspace{-0.05in}
    \label{tab:variations1}
	\centering
	\begin{tabular}{lp{2in}}
	    \toprule
        \textbf{Variations} & \textbf{Description}   \\
        \midrule
        \sysdiff & differential comparison version \\
        \hspace{0.2in} \sysdiff-w/ ({\it default}) & \sysdiff with generated non-member set  \\
       \hspace{0.2in} \sysdiff-w/o & \sysdiff without generated non-member set  \\
       \midrule
       \syssvm & one-class SVM version with generated non-members as training set\\
        \bottomrule
	\end{tabular} \vspace{-0.1in}
\end{table}

We now describe the overall procedure of \sys in Figure~\ref{fig:arch}.  \sys takes target samples with unknown membership and outputs a membership result for each individual sample.  Specifically, \sys first generates a non-member dataset and then queries the target DNN model with the target and the non-member datasets to obtain the output probabilities.  Then, \sys applies a projection function to select certain important features from the output probabilities.  The next step depends on different variations of \sys.  \sysdiff adopts differential comparison to classify members and non-members
in the target dataset; \syssvm trains a one-class model from the selected output probabilities of the non-member and classifies samples in the target dataset using the trained model.  We list different variations of \sys in Table~\ref{tab:variations1}. \textcolor{blue}{Both the \sysdiff-w/ and \syssvm require a generated nonmember set as opposed to \sysdiff-w/o that does not require it.  The default \sys is \sysdiff-w/. We introduce \sysdiff-w/o since it may be hard to create an effective nonmember set in some cases and the created set could be contaminated; at the same time, we introduce \syssvm more as a baseline to separate the effectiveness of differential comparison and the generation of a nonmember set.}

 
 \textcolor{blue}{Next, we introduce two different modes of \sys: batch and incremental.  The batch mode takes and classifies a batch of target samples iteratively and the incremental mode takes one sample, adds to a previous batch, and then classifies the target.  It is worth noting that both modes give the same inference results for given samples. They are designed to handle different scenarios: The batch mode for a cluster of samples and the incremental mode for each individual sample. }

\subsection{Dataset Preparation for Differential Comparison} \label{Nonmember_Generation}

 \def\semichecked{\ding{51}\!\!\!\!\!\raisebox{0.2 em}{$\smallsetminus$}} 
\begin{table}[!t]
 \renewcommand{\arraystretch}{1.2}
 \setlength{\tabcolsep}{2pt}
 \scriptsize
    \caption{Methods in Generating Non-members} \vspace{-0.1in}
    \label{tab:nonmembers}
	\centering
	\begin{tabular}{lccc}
	    \toprule
        \textbf{Options} & \textbf{Description} \\
        \midrule
        Sample transform & Laplace, Sobel, Scharr, and Canny \\
         Random perpetuation  & Gaussian or Salt Pepper noise \\ 
       Random generation & Random feature values \\
       Cross domain & Samples from a different domain \\
        \bottomrule
	\end{tabular} \vspace{-0.15in}
\end{table}

In this subsection, we describe how to generate non-members for a target model. There are two general directions of non-member set generation: (i) generating new samples or transforming existing ones and (ii) roughly separating a set with existing samples into two with members and non-members. 

\subsubsection{Generation of Non-members}
 Let us start from the first direction.  This is applicable if an adversary can probe the target model with arbitrary samples.  From a high-level perspective, because the input space is usually much larger than the training set, the adversary can generate a new sample, which is likely not in the training set. 
 We now discuss four generation methods below and in Table~\ref{tab:nonmembers}.

\begin{icompact}
\item Sample Transformation.  \hspace{0.01in} An adversary applies an operator, e.g., Laplace, Sobel, Scharr, and Canny, on an existing sample to obtain a new one. Take Sobel for example: the adversary transforms an image to one emphasising edges.  The advantage of this method is that it usually preserves some semantics, thus being effective to be distinguishable from members.  Additionally, the generated sample is stealthy as all the operators are commonly used in image processing. The down-side of this method is that many operators are specific to the image domain. 
\item Random Perpetuation. \hspace{0.01in} An adversary adds random noises, like Gaussian and Salt Pepper,  to an existing sample for the generation. This method is also effective as many semantics are preserved, but less stealthy because one may detect noise levels in the frequency domain. 
\item Random Generation. \hspace{0.01in} An adversary generates a sample with random features.  This method is less effective as the generated samples, e.g., a random image, may not have any semantics, and less stealthy as any human can easily spot the generated sample as a noise. 
\item Cross-domain Samples.\hspace{0.01in} An adversary may adopt samples from another domain, e.g., a celebrity face dataset for a model trained with CIFAR-100.  This is also effective but less stealthy, because the probed samples are apparently from a different domain. 
\end{icompact}

Note that samples generated following Table~\ref{tab:nonmembers} are highly likely non-members. Therefore, these samples can be used in both \syssvm and \sysdiff, particularly the training set of  \syssvm and the comparison set of \sysdiff. 

\subsubsection{Rough Sample Separation}
We then describe the second direction. This is applicable when the adversary does not have free probing access to the target model, but only obtain the output probability distribution of a limited dataset.  This scenario may happen if the adversary is only allowed to probe samples from a certain source, e.g., disease images acquired at a specific hospital.  
 There are two different methods used in such separation: (i) a clustering algorithm like k-means and \textcolor{blue}{agglomerative clustering}, and (ii) a separation based on the highest probability score. The first method  is to apply clustering to roughly divide the target dataset into two, one as members and the other as non-members. The second is to roughly select those with high probability score as members and those with low as non-members. 
%
 This generation is only applicable to \sysdiff, because the non-member set may be \highlight{noisy}. 

\subsection{Probability Score Projection}

We now discuss our probability score projection function $G_{projection,k}$, which applies on $S^{prob}_{target}$ to obtain $k$ different elements.  The high-level idea is that class types, e.g., a bird vs. a tree, are less important features for an MI attack, but the ranking of values in different classes determines the membership. 
 Based on this insight, we design three different \highlight{projection} functions.



\begin{icompact}
\item All probability scores in an order.  \hspace{0.05in} This \highlight{projection} function ranks all probability scores from the largest to the smallest, which removes the class information but only keeps the relative values. 
\item Top-k probability scores.   \hspace{0.05in} This \highlight{projection} function selects the top-k probability scores to further remove some noisy ones with small values. 
\item Top-k + Ground Truth Class. \hspace{0.05in} This \highlight{projection} function---used in the blackbox setting---further includes the value corresponding to the ground truth class. 
\end{icompact}

\subsection{Differential Comparison}

In this section, we describe one key technique, i.e., differential comparison, in this paper. The first task is to calculate the distance between two sets. Just like all general ML tasks, it is hard to differentiate member and non-members directly in the output probability distribution space. Therefore, \sys maps all probabilities to the Reproducing Kernel Hilbert Space (RKHS)~\cite{MMD} and then calculates the distance between two centroids in the kernel space. Specifically, our distance, based on Maximum Mean Discrepancy (MMD)~\cite{MMD1}, is shown in Equation~\ref{equ-4}.
\begin{equation} \footnotesize
D(S^{prob,k}_{target}, S^{prob,k}_{nonmem}) = \left \| \frac{1}{n_{t}}\sum_{i = 1}^{n_{t}}\phi(y_i)- \frac{1}{n_{n}}\sum_{j = 1}^{n_{n}}\phi(y^\prime_j)\right \|_\nu 
\label{equ-4}
\end{equation} where $y_i \in S^{prob,k}_{target}$, $y^\prime_j \in S^{prob,k}_{nonmem}$, $n_t$ and $n_n$ are the size of $S^{prob,k}_{target}$ and $S^{prob,k}_{nonmem}$, $\nu$ is the dimension of the kernel space, and $\phi$ is a feature space map $k \mapsto \nu$, \highlight{e.g.,} a Gaussian kernel function $k(y, y^\prime) = \langle\phi(y), \phi(y^\prime)\rangle= exp(-||y-y^\prime||/(2\sigma^2))$.


The second task is to perform differential comparison between two sets. There are two variations of differential comparison, single- and bi-directional, which defines the direction in moving samples between two sets.



\subsubsection{Single-directional Differential Comparison}

This method 
 iteratively moves samples from $S^{prob,k}_{target}$ to $S^{prob,k}_{nonmem}$, compares the distance before and after move, and then determines the moved sample's membership.  Details of the method are shown in Algorithm~\ref{algo-1}. Lines 1--2 of Algorithm~\ref{algo-1} prepare some initial variables and then Lines 3--14 are the iterative algorithm.  Specifically, Line 5 first calculates the distance between two sets and Lines 6--12 go through all the elements in $S^{prob,k}_{target}$.  If the updated distance after moving a sample (Line 7) is larger than the original (Line 8), \sysdiff considers it as a non-member. After one iteration, \sysdiff updates $S^{prob,k}_{target}$ (Line 13) and starts the entire process again.

 

\renewcommand{\algorithmicrequire}{\textbf{Input:}}
\renewcommand{\algorithmicensure}{\textbf{Output:}}
    

    \begin{algorithm}[!t] \footnotesize
        \caption{Single-directional Differential Comparison}
        \label{algo-1}
        \begin{algorithmic}[1] 
            \Require $S^{prob,k}_{nonmem}$, $S^{prob,k}_{target}$ 
            \Ensure $S_{pred,nonmem}$, $S_{pred,mem}$
                \State $S_{pred,nonmem} \gets empty$
                 \State $flag \gets true$
                \While{flag}
                   \State $flag \gets false$
                    \State $ d \gets D(S^{prob,k}_{nonmem}, S^{prob,k}_{target})$
                    \For{$y \in S^{prob,k}_{target}$}
                        \State $d' \gets D(S^{prob,k}_{nonmem} \cup \{y\}, S^{prob,k}_{target} - \{y\})$
                        \If{$d'\geq d$}
                            \State $S_{pred,nonmem} \gets S_{pred,nonmem} \cup {y}$
                             \State $flag \gets true$
                        \EndIf
                    \EndFor
                    \State $S^{prob,k}_{target} \gets S^{prob,k}_{target} - S_{pred,nonmem} $
                \EndWhile
                \State $S_{pred,mem} \gets S^{prob,k}_{target}$
        \end{algorithmic}
    
    \end{algorithm}

\subsubsection{Bi-directional Differential Comparison}

    
        \begin{algorithm}[!t] \footnotesize
        \caption{Bi-directional Differential Comparison}
        \label{algo-2}
        \begin{algorithmic}[1] 
            \Require $S^{prob,k}_{target1}$, $S^{prob,k}_{target2}$ 
            \Ensure $S_{pred,nonmem}$, $S_{pred,mem}$
                \State $flag \gets true$
                \While{flag}
                  \State $flag \gets false$
                    \State $ d \gets D(S^{prob,k}_{target1}, S^{prob,k}_{target2})$
                    \For{$y \in S^{prob,k}_{target1}$}
                        \State $ d' \gets D(S^{prob,k}_{target2} \cup \{y\}, S^{prob,k}_{target1} - \{y\})$
                        \If{$d'\geq d$}
                            \State $S^{prob,k}_{target1} \gets S^{prob,k}_{target1} - \{y\}$
                            \State $S^{prob,k}_{target2} \gets S^{prob,k}_{target2} \cup \{y\}$
                            \State $flag \gets true$
                            \State $d \gets d'$
                        \EndIf
                    \EndFor
                    \For{$y \in S^{prob,k}_{target2}$}
                        \State $ d' \gets D(S^{prob,k}_{target1} \cup \{y\}, S^{prob,k}_{target2} - \{y\})$
                        \If{$d'\geq d$}
                            \State $S^{prob,k}_{target2} \gets S^{prob,k}_{target2} - \{y\}$
                            \State $S^{prob,k}_{target1} \gets S^{prob,k}_{target1} \cup \{y\}$
                            \State $flag \gets true$
                            \State $d \gets d'$
                        \EndIf
                    \EndFor
                \EndWhile
                \State $S_{pred,mem}, S_{pred,nonmem} \gets S^{prob,k}_{target1}, S^{prob,k}_{target2}$
        \end{algorithmic}
    
    \end{algorithm}

This method works on a roughly divided two datasets, say  $S^{prob,k}_{target1}$ and $S^{prob,k}_{target2}$, and moves samples in both directions, i.e., $S^{prob,k}_{target1} \rightarrow S^{prob,k}_{target2}$ and $S^{prob,k}_{target2} \rightarrow S^{prob,k}_{target1}$.  More specifically, the method details are shown in Algorithm~\ref{algo-2}.  \sysdiff first moves samples from $S^{prob,k}_{target1}$ to $S^{prob,k}_{target2}$ in Lines 5--13, and then $S^{prob,k}_{target2}$ to $S^{prob,k}_{target1}$ in Lines 14--22.  Then, \sysdiff iterates the entire procedure until it converges.

Note that one major challenge here is to decide whether $S^{prob,k}_{target1}$ or $S^{prob,k}_{target2}$ contains
non-members, as those two sets are symmetric and look the same.  The intuition here is that the average prediction confidence score of members is higher than the one of non-members.  Therefore,  \sysdiff compares the average confidence score for a decision in the end.

\subsection{Batch Division and Size Optimization}

In this part, we discuss how \sys divides the target dataset into small batches with an appropriate size especially when the size of nonmember dataset is small. The high-level idea of determining the size is that \sys needs to maximize the distance change in differential comparison when moving one sample. Specifically, \sys starts from a batch size consistent with the size of nonmember dataset. Such an algorithm keeps \sys sensitive while still maintaining a small size of non-members.  

\section{Datasets, Prior Attacks and Implementation}\label{Implementation and Dataset}

In this section, we describe the datasets used in the experiments, target and shadow models, existing state-of-the-art attacks, and our implementation of \sys. 

\subsection{Datasets} \label{dataset}

\begin{table*}[!t]
   \renewcommand{\arraystretch}{1.2} \setlength{\tabcolsep}{2pt}
   \scriptsize
    \caption{A description of different datasets used in the evaluation. }
    \label{tab:dataset}
	\centering
	\begin{tabular}{cclccccc}
		\toprule
	    \textbf{Dataset}  &\textbf{\# of classes} & \textbf{Description} & \textbf{Resolution} & \textbf{\# Epochs (target model)} & \textbf{Training set (target model)} &  \textbf{Training set (shadow model)} & \textbf{Target set}  \cr
        \midrule
        \textcolor{blue}{Adult} &  2  & census income records  & N/A  & 100 & 16,280 & 16,280 &32,560  \cr
        EyePACS & 5 & retina images with diabetic retinopathy  & 150$\times$150 & (pre-trained + 15) or 150 & 10,000 & 10,000 & 20,000 \cr
        CH-MNIST& 8 & histological images of colorectal cancer & 64$\times$64 & (pre-trained + 15) or 150 & 2,500 & 2,500 & 5,000\cr
        \textcolor{blue}{Location} &  30  & mobile users' location check-in records  & N/A  & 100 & 2,505 &  2,505& 5,010 \cr
        Purchase-50 &  50  & shoppers' purchase histories & N/A  & 100 & 10,000 &  10,000& 20,000 \cr
        \textcolor{blue}{Texas} &  100  & inpatients stays in health facilities& N/A  & 100 & 10,000 &  10,000& 20,000 \cr
        CIFAR-100& 100 &  object recognition dataset & 32$\times$32 & (pre-trained + 30) or 150 & 10,000 & 10,000 & 20,000\cr
        Birds-200& 200 & photos of birds species & 150$\times$150 & (pre-trained + 15) or 150 & 5,894 & 5,894 & 11,788 \cr
        \bottomrule
	\end{tabular}
\end{table*}

We use eight datasets as shown in Table~\ref{tab:dataset} to evaluate \sys on different application scenarios.

\subsubsection{UCI Adult}

\textcolor{blue}{UCI Adult, or Adult for short, has 48,842 records with census attributes, such as age, gender, education, marital status, and working hours. The classification task is to predict whether a person earns over \$50,000 per year based on given attributes. We follow a well-known preprocessing method\footnote{\url{https://github.com/rupampatir/TrainingDataSynthesizer/blob/master/classifiers/income/income_classifier.py}} to obtain a target datasets with 32,560 records---half are used as the training set of the target model and half as the training set of the shadow model. The target dataset contains all the samples.}

\subsubsection{EyePACS}

The EyePACS dataset  from 
  Kaggle's  was used for a Diabetic Retinopathy Detection challenge\footnote{\url{https://www.kaggle.com/c/diabetic-retinopathy-detection/data}}. The dataset includes 88,703 high-resolution retina images taken under a variety of imaging conditions and each image has a label ranging from 0 to 4,  representing the presence and severity of diabetic retinopathy. We adopt the preprocessing method from Kaggle\footnote{\url{https://www.kaggle.com/ratthachat/aptos-eye-preprocessing-in-diabetic-retinopathy}}. We select 10,000 random images as the training set of the target model, 10,000 disjoint images as the training set of the shadow model, and 20,000 images---i.e., 10,000 members and 10,000 non-members---as the target set for inference.


\subsubsection{CH-MNIST} CH-MNIST~\cite{CH-MNIST} is a benchmark dataset of 5,000 histological images of human colorectal cancer including 8 classes of tissues.  We obtain a version\footnote{\url{https://www.tensorflow.org/datasets/catalog/colorectal_histology}} of CH-MNIST from TensorFlow Datasets, in which each image's resolution is 150$\times$150. We resize all  images to 64$\times$64 to increase the diversity of image resolution, and then randomly select two sets of 2,500 images as training data of the target and the shadow models. The target dataset has all 5,000 images, i.e., 2,500 members and 2,500 non-members for the target model.  Note that due to the small size of CH-MNIST, the training sets of shadow and target models have overlap.

\subsubsection{Location}
\textcolor{blue}{This dataset is from the publicly available set of mobile users' location ``check-ins'' in the Foursquare social network\footnote{\url{https://sites.google.com/site/yangdingqi/home/foursquare-dataset}}. We obtain a processed version of the dataset from a prior work~\cite{earliest}, which has 5,010 record with 446 binary features and is clustered into 30 classes, each representing a different geosocial type. The task is to predict the user's geosocial type given his or her record. We use the whole dataset and randomly chose samples to create two sets, each with 2,505 samples, to train the target model and the shadow model respectively. There are overlapping samples in both target and shadow models' training sets since the dataset is small.}

\subsubsection{Purchase-50} Purchase-50 dataset is from Kaggle's ``Acquired Valued Shoppers Challenge''\footnote{\url{https://www.kaggle.com/c/acquire-valued-shoppers-challenge/data}} and contains purchase histories of many shoppers. We obtain a simplified version with 197,324 records from R.Shokri et al.~\cite{earliest}, where each record contains 600 binary features representing whether the customer has purchased an item. We cluster the dataset into 50 classes, in which each class represented a different purchase habit. The training datasets of target and shadow models are disjoint with 10,000 samples each; The target dataset has 20,000 samples, i.e., 10,000 members and {10,000} nonmembers.


%

\subsubsection{Texas hospital stays}

\textcolor{blue}{Texas hospital stays, or Texas for short, is the inpatient stays records in several health facilities based on the Hospital Discharge Data released by Texas Department of State Health Services from 2006 to 2009. We follow the same preprocessing method and classification task as prior work~\cite{earliest}.  The training datasets of target and shadow models are disjoint with 10,000 samples each; We also select 20,000 records for the target dataset, i.e., {10,000} members and {10,000} nonmembers. }

\subsubsection{CIFAR-100} CIFAR-100 is a popular benchmark dataset that is used to evaluate image recognition algorithms.  The dataset has 60,000 images evenly distributed over 100 classes. We randomly select two sets of 10,000 images evenly distributed over 100 classes as the training datasets of the target model, and another disjoint 10,000 images as the training datasets of the shadow model.  The target dataset has {20,000} images: {10,000} members and {10,000} nonmembers. 


\subsubsection{Caltech-UCSD Birds 200} Caltech-UCSD Birds 200~\cite{WelinderEtal2010}, or for short Birds-200, is an image dataset with photos of mostly North American birds species.  The dataset has 11,788 images from 200 classes.  In our experiments the training dataset of target and shadow models each has 5,894 samples; The target dataset has {5,894} members and {5,894} nonmembers


\subsection{Target and Shadow Models}
%

\begin{table}[!t]
    \renewcommand{\arraystretch}{1.2} \scriptsize \setlength{\tabcolsep}{2.5pt}
    \caption{Target and shadow models' architecture and hyper-parameter setting}
    \label{tab:modelarch}
	\centering
	\begin{threeparttable}
	\begin{tabular}{cccccc}
		\toprule
		\multirow{2}{*}{\textbf{Model arch.}} & \multirow{2}{*}{\textbf{\# of layers}}  & \multicolumn{2}{c}{\bf Target model } & \multicolumn{2}{c}{\bf Shadow model (blackbox)} \cr 
		\cmidrule(l{2pt}r{2pt}){3-4} \cmidrule(l{2pt}r{2pt}){5-6}
	     && \textbf{Max. epochs} &  \textbf{LRN rate} & \textbf{Max. epochs} &  \textbf{LRN rate}  \cr
        \midrule
        ResNet50 & 50 & p$^*$+m$^{**}$ & $5e^{-5}$& p+0.2m & $5e^{-5}$\cr
        ResNet101 & 101 & p+m & $5e^{-5}$& p+0.3m & $1e^{-4}$\cr
        VGG16& 16 & p+m & $5e^{-5}$& p+0.6m & $5e^{-5}$\cr
        DenseNet121 &  121 & p+m & $5e^{-5}$& p+m & $1e^{-4}$\cr
        VGG19&  19 & p+m & $5e^{-5}$& p+1.5m & $5e^{-5}$\cr
        Standard CNN& 2 & m & $5e^{-5}$&  0.5m  &$1e^{-4}$\cr
        MLP & [3--7] dense & m & $5e^{-5}$&  [0.3--2]m &$1e^{-4}$/$5e^{-5}$\cr
        \bottomrule
	\end{tabular}
	\begin{tablenotes}
        \footnotesize
        \item[*]p: the epoch of a pre-trained weight on the ImageNet dataset; 
        \item[**]m: the maximum epoch of target model for each dataset in Table~\ref{tab:dataset}.
      \end{tablenotes}
    \end{threeparttable}
\end{table}

In this part, we describe the architectures and hyper-parameters of target and shadow models of our evaluation in Table~\ref{tab:modelarch}.  We adopt seven different popular DNN architectures with pre-set maximum epochs and learning rate.  Note that all popular DNNs, e.g., ResNet, VGG, and DenseNet, are the standard architectures with pre-trained parameters from ImageNet; we adopt the same standard CNN architecture and hyperparameters as prior blackbox MI attack~\cite{earliest}; the multilayer perceptron (MLP) model has at most seven dense layers with size of 8192, 4096, 2048, 1024, 512, 256, and 128 and an additional Softmax layer. 
 Now we describe how we select and train target and shadow models. 

\begin{icompact}
\item Target model. \hspace{0.05in} Given a dataset, we randomly select a model architecture from the target model column of Table~\ref{tab:modelarch} and train the model with the specified hyperparameters. 
\item Shadow model (blackbox and blind settings).   \hspace{0.05in} \highlight{Given} a target model and a dataset, we randomly select and train a model with the architecture and hyperparameters specified in the shadow model column of Table~\ref{tab:modelarch}.
\item Shadow model (graybox and graybox-blind settings). \hspace{0.05in} \highlight{Given} a target model and a dataset, we select the same architecture and hyperparameters as the target model. 
\end{icompact}

\subsection{State-of-the-art Attacks} \label{subsec:stateoftheartattacks}

\begin{table}[!t]
    \renewcommand{\arraystretch}{1.2}
      \scriptsize \setlength{\tabcolsep}{2.5pt}
    \caption{A list of conditions of baseline MI attacks and different variations of \sys}
    \label{tab:attacks}
	\centering
	\begin{threeparttable}
	\begin{tabular}{ccccc}
		\toprule
	     \textbf{Attacks}  &\textbf{True labels} & \textbf{Shadow} &  \textbf{Threat model} & \textbf{Target Model Probes} \cr
        \midrule
        NN~\cite{earliest} & \ding{55} & \ding{51} & all & Target set \cr
        Top3-NN~\cite{NDSS} & \ding{55} & \ding{51} & all & Target set\cr
        Top1-Thre~\cite{NDSS}& \ding{55} & \ding{55} & all & Target set + 1,000 samples\cr
        Loss-Thre~\cite{globalloss}&  \ding{51} & \ding{51} & blk, gray & Target set\cr
        Label-Only~\cite{globalloss}&   \ding{51} & \ding{55} & blk, gray & Target set\cr
        Top2+True& \ding{51} & \ding{51} & blk, gray & Target set \cr
        \midrule
        \sysdiff-w/ & \ding{55}\ding{51} & \ding{55} & all & Target set + 20 samples \cr
        \sysdiff-w/o & \ding{55}\ding{51} & \ding{55} & all & Target set \cr
        \syssvm & \ding{55}\ding{51} & \ding{55} & all & Target set + 1,000 samples \cr
        \bottomrule
	\end{tabular}
	\begin{tablenotes}
        \footnotesize
        \item[*]\ding{55}\ding{51}: The approach works either with or without the condition, e.g., the ground truth labels.
      \end{tablenotes}
    \end{threeparttable}
\end{table}

In this part, we describe state-of-the-art attacks in the literature as shown in Table~\ref{tab:attacks}.  We follow the descriptions in prior work to implement each attack for the comparison with \sys. Generally speaking, there are two categories, \highlight{those} without ground-truth labels and \highlight{those} with ground-truth labels.

\subsubsection{Attacks without Ground-truth Labels} \label{attackwithoutlabel}

We describe three prior attacks without ground-truth labels. 
 Presumably, those attacks work under all settings, but their performance are the same, with or without ground-truth label information, i.e., under blind and blackbox settings. 

\begin{icompact}
\item {Neural network (NN).} \hspace{0.05in} The NN-based MI attack
\textcolor{blue}{trains a NN from all features from the output probability distributions of a shadow model. We follow both Shokri et al.~\cite{earliest} and Salem et al.~\cite{NDSS} for the implementation. 
 } 
\item {Neural  network with top three features (Top3-NN).} \hspace{0.05in} This MI attack proposed by Salem et al.~\cite{NDSS}
, which trains an NN based on the top three features from the output probability distributions of a shadow model. 
\item {Threshold based on top one feature (Top1-Thre).} \hspace{0.05in} This MI attack, which is also proposed by Salem et al.~\cite{NDSS} as their Adversary Three, compares the top feature from the output probability distribution with a threshold and classifies the sample as member if the top feature is larger than the threshold. 
\end{icompact}

\subsubsection{Attacks with Ground-truth Labels}

We describe three attacks that specifically require ground-truth labels: They may or may not need a shadow model.  That is, these attacks only work under settings where ground-truth labels are available, i.e., blackbox and graybox settings.


\begin{icompact}
\item {Threshold based on a loss function (Loss-Thre).} \hspace{0.05in} This MI attack from Yeom et al.~\cite{globalloss}, which requires a shadow model, computes a cross-entropy loss, $loss=-log(F_{T}(x)_y)$, where $F_{T}(x)_y$ is the probability of the true label $y$ of the data sample $x$, and classifies $x$ as a member if $loss$ is smaller than the average loss of all training samples in the shadow model.


\item {Discrepancy between predicted and ground-truth class (Label-Only).} \hspace{0.05in} This MI attack from Yeom et al.~\cite{globalloss}, which does not requires a shadow model, classifies a sample as a member if the predicted class is the same as the ground-truth one. 
\item {Neural network with top two feature plus the feature with \highlight{ground-truth} label (Top2+True).} \hspace{0.05in} This MI attack is an improved version 
of the NN attack from Shokri et al.~\cite{earliest} and Salem et al.~\cite{NDSS} with the consideration of the ground-truth label.  We add this attack as a baseline for the comparison purpose. 
\end{icompact}

\subsection{Implemenatation} 



We implemented \sys with 811 lines of code (LoC) based on TensorFlow 2.1.0. 
  Specifically, our implementations of \syssvm, \sysdiff-w/, and \sysdiff-w/o are of 227, 261 and 323 Lines of Python 3.7 code respectively. The non-member generation module has 72 LoC, the differential comparison module 182 LoC. We also implement prior attacks with 344 LoC.  Our implementations of \sys and prior attacks are open-source and available at this anonymous repository: \url{https://github.com/hyhmia/BlindMI}. 

\section{Evaluation} \label{Evaluation}

We first introduce the evaluation metrics and several research questions (RQs). Then, we show the performances of MI attacks under different settings based on different RQs, and explain what we learn from the results in details.

\subsection{Evaluation Metrics, Experimental Setting and Research Questions} \label{subsec:expsetting}

We mainly use F1-score, the harmonic mean of precision and recall, as our evaluation metrics, because F1-score represents a trade-off between precision and recall. Specifically, Precision represents the ratio of real-true members predicted among all the positive membership predictions made by an adversary, and Recall demonstrates the ratio of true members predicted by an adversary among all the real-true members.
 \highlight{We adopt the batch mode for \sys in our experiments. Following the prior work~\cite{NDSS}, in the Blind and Graybox-Blind settings, we select the top three feature values for all variations of \sys; in the Blackbox and Graybox settings, we select the top two feature values plus the value of the ground-truth class. } \textcolor{blue}{All the experiments are performed using the GeForce RTX 2080 graphics cards (NVIDIA).}
 


Our evaluation aims to answer the following RQs. 

\begin{icompact}

\item {\it RQ1 [All Settings]:} What is the performance of all variations of \sys compared with state-of-the-art MI attacks under different settings?
\item {\it RQ2 [Blackbox Setting]:} How does \sys perform under existing defenses against MI attacks?
\item {\it RQ3 [Blind Setting]:} What is the performance of \sys for different quality and size of the non-member set? 
\item \textcolor{blue}{{\it RQ4 [Blind Setting]:} How do different initial classifiers and kernel functions affect the performance of \sysdiff?}
\item \textcolor{blue}{{\it RQ5 [Blind Setting]:} How long and how many moves and iterations are needed for \sysdiff to converge?}
\item {\it RQ6 [Blackbox Setting]:} What is the performance of \sys under different real-world settings, e.g., nonmember-to-member ratio and number of target model's classes?
\end{icompact}

%
%
%
%
%
%
%
%

\subsection{RQ1: Attack Performance With Different Settings} \label{RQ1}

In this subsection, we evaluate and compare the \textcolor{blue}{Precision, Recall, and} F1-score of \sys and existing attacks in Section~\ref{subsec:stateoftheartattacks}. Our setting for this RQ is that the nonmember dataset size of \sysdiff-w/ is 20, the nonmember dataset size of \syssvm is 1,000, and \sysdiff-w/o does not need additional nonmembers. The target dataset sizes depending on the problem domain are shown in Table~\ref{tab:dataset}. Each attack is performed ten times with a new target and shadow model with different training datasets, model architectures and hyperparameters each time. Then, we obtain the average values of F1-score together with the standard error of the mean among the ten attacks. 

\begin{table*}[!t]
\setlength{\tabcolsep}{3pt}
   \renewcommand{\arraystretch}{1.05}
   \scriptsize
    \caption{\textcolor{blue}{Precision, Recall, F1-Score (\%) with standard error of the mean of prior attacks and \sys under four different adversarial settings (\textcolor{blue1}{Blue} indicates the highest recall, \textcolor{og}{green} the highest precision, and \textcolor{red}{red} the highest F1-score for each setting).}} 
    \label{table_results}
	\centering
	\begin{tabular}{M{3mm}|M{20mm}|M{12mm}M{15mm}M{15mm}M{15mm}M{15mm}M{15mm}M{15mm}M{15mm}M{15mm}}
	    \toprule
        &\textbf{Attack} &\textbf{Metric} & \textbf{Adult} &\textbf{EyePACS} & \textbf{CH-MNIST} & \textbf{Location} & \textbf{Purchase-50} &  \textbf{Texas}& \textbf{CIFAR-100} & \textbf{Birds-200} \cr
        \midrule
        \midrule
        \multirow{18}{*}{\rotatebox[origin=c]{90}{\textbf{Blind}}} & \multirow{3}{*}{NN} & Precision & 49.9 $\pm$ 0.30 & 56.6 $\pm$ 3.80 & 62.5 $\pm$ 0.62 & 68.9 $\pm$ 5.09 & 61.0 $\pm$ 1.64 & 62.5 $\pm$ 3.02 & 73.2 $\pm$ 12.1 & 95.1 $\pm$ 1.90 \cr
        & & Recall & 35.1 $\pm$ 10.1 & 90.5 $\pm$ 9.58 & 85.8 $\pm$ 8.22 & 93.6 $\pm$ 6.39 & 67.1 $\pm$ 19.9 & 99.8 $\pm$ 0.15 & 99.5 $\pm$ 0.28 & 53.5 $\pm$ 27.5 \cr
        & & F1-Score & 40.6 $\pm$ 7.32 & 69.1 $\pm$ 0.02 & 71.7 $\pm$ 3.53 & 78.4 $\pm$ 3.23 & 59.4 $\pm$ 11.9 & 76.7 $\pm$ 2.20 & 83.1 $\pm$ 3.53 & 58.3 $\pm$ 27.4 \cr
        \cline{3-11}

        & \multirow{3}{*}{Top3-NN} & Precision & 49.8 $\pm$ 0.37 & 55.7 $\pm$ 3.69 & 62.1 $\pm$ 0.19 & 69.1 $\pm$ 4.98 & 61.0 $\pm$ 1.73 & 62.5 $\pm$ 2.80 & 70.3 $\pm$  9.33 & 98.1 $\pm$ 0.08 \cr
        & & Recall & 20.2 $\pm$ 7.34 & 93.3 $\pm$ 6.58 & 84.8 $\pm$ 9.70  & 92.8 $\pm$ 7.17 & 67.6 $\pm$ 19.9 &\textcolor{blue1}{\textbf{99.9 $\pm$ 0.08}} & \textcolor{blue1}{\textbf{99.6 $\pm$ 0.11}} & 60.3 $\pm$ 23.1 \cr
        & & F1-Score & 26.7 $\pm$ 7.25 & 69.5 $\pm$ 1.04 & 70.9 $\pm$ 4.03 & 78.1 $\pm$ 3.39 & 59.6 $\pm$ 12.1 & 76.8 $\pm$ 2.07 & 81.7 $\pm$ 6.66 & 68.6 $\pm$ 21.3 \cr
        \cline{3-11}
        
        & \multirow{3}{*}{Top1-Threshold} & Precision & 0.51 $\pm$ 0.23 & \textcolor{og}{\textbf{99.9 $\pm$ 0.03}}& \textcolor{og}{\textbf{74.7 $\pm$ 24.9}} & 14.1 $\pm$ 2.93 & 61.6 $\pm$ 12.1 & 0.34 $\pm$ 0.20 & \textcolor{og}{\textbf{97.2 $\pm$ 1.70}} & \textcolor{og}{\textbf{99.9 $\pm$ 0.06}} \cr
        & & Recall & 31.3 $\pm$ 13.6 & 55.2 $\pm$ 0.51 & 40.9 $\pm$ 13.7 & 63.7 $\pm$ 1.60 & 63.2 $\pm$ 1.97 & 24.6 $\pm$ 14.2 & 86.2 $\pm$ 3.55 & 57.9 $\pm$ 0.81\cr
        & & F1-Score & 1.01 $\pm$ 0.44 & 71.1 $\pm $ 0.42 & 52.8 $\pm$ 17.6 & 22.7 $\pm$ 3.87 & 53.5 $\pm$ 7.26 & 0.67 $\pm$ 0.38 & 92.8 $\pm$ 1.72 & 71.4 $\pm$ 0.65 \cr
        \cline{2-11}

        & \multirow{3}{*}{\textbf{BlindMI-Diff-w/}} & Precision & 50.0 $\pm$ 0.03 & 66.2 $\pm$ 2.80 & 65.4 $\pm$ 2.52 & 75.9 $\pm$ 1.44 & \textcolor{og}{\textbf{66.0 $\pm$ 0.55}} & \textcolor{og}{\textbf{76.1 $\pm$ 1.06}} & 90.7 $\pm$ 0.86 & 95.5 $\pm$ 0.08 \cr
        & & Recall & \textcolor{blue1}{\textbf{90.4 $\pm$ 5.95}}& 94.4 $\pm$ 3.35 & 88.8 $\pm$ 2.04 & 99.9 $\pm$ 0.08 & 95.2 $\pm$ 1.38 & 97.4 $\pm$ 1.64 & 97.4 $\pm$ 0.37 & 98.2 $\pm$ 0.26\cr
        & & F1-Score & \textcolor{red}{\textbf{64.2 $\pm$ 1.59}}  & \textcolor{red}{\textbf{77.7} $\pm$ \textbf{0.80}}  & \textcolor{red}{\textbf{75.1} $\pm$ \textbf{1.49}} &  \textcolor{red}{\textbf{86.2} $\pm$ \textbf{0.90}} & \textcolor{red}{\textbf{78.0} $\pm$ \textbf{0.31}} & \textcolor{red}{\textbf{85.5 $\pm$ 0.80}} & \textcolor{red}{\textbf{93.9} $\pm$ \textbf{0.63}} & \textcolor{red}{\textbf{96.8} $\pm$ \textbf{0.09}} \cr
        \cline{3-11}
        
        & \multirow{3}{*}{\textbf{BlindMI-Diff-w/o}} & Precision & \textcolor{og}{\textbf{50.5 $\pm$ 0.05}} & 60.3 $\pm$ 1.26 & 63.3 $\pm$ 1.60 & \textcolor{og}{\textbf{71.5 $\pm$ 0.87}} & 61.7 $\pm$ 0.38 & 69.2 $\pm$ 4.09 & 86.1 $\pm$ 0.78 & 94.5 $\pm$ 0.24 \cr 
        & & Recall & 84.2 $\pm$ 4.05 & \textcolor{blue1}{\textbf{99.3 $\pm$ 0.10}} &  92.4 $\pm$ 2.68 & \textcolor{blue1}{\textbf{99.9 $\pm$ 0.04}} & \textcolor{blue1}{\textbf{99.5 $\pm$ 0.25}}& 97.6 $\pm$ 1.83& 98.9 $\pm$ 0.60 & 98.0 $\pm$ 0.80 \cr
        & & F1-Score & 62.7 $\pm$ 1.12 & 75.0 $\pm$ 1.40 & 75.1 $\pm$ 1.89 & 83.3 $\pm$ 0.57 & 76.2 $\pm$ 0.25 & 80.7 $\pm$ 2.37 & 92.1 $\pm$ 1.15 & 96.2 $\pm$ 0.26 \cr
        \cline{3-11}
        
        & \multirow{3}{*}{\textbf{BlindMI-1Class}} & Precision & 49.9 $\pm$ 0.07 & 64.7 $\pm$ 2.41 & 59.0 $\pm$ 1.13 & 63.4 $\pm$ 0.44 & 61.1 $\pm$ 1.38 & 70.9 $\pm$ 1.11 & 83.5 $\pm$ 0.40 & 83.8 $\pm$ 11.9 \cr
        & & Recall & 56.5 $\pm$ 5.85 & 94.4 $\pm$ 0.01 & \textcolor{blue1}{\textbf{94.8 $\pm$ 1.41}} & 99.7 $\pm$ 0.27 & 98.1 $\pm$ 0.71 & 97.2 $\pm$ 2.07 & 98.9 $\pm$ 0.69 & \textcolor{blue1}{\textbf{99.1 $\pm$ 0.60}} \cr
        & & F1-Score & 52.6 $\pm$ 2.47 & 76.8 $\pm$ 1.70 & 72.8 $\pm$ 1.27 & 77.6 $\pm$ 0.33 & 77.1 $\pm$ 0.50 & 81.9 $\pm$ 0.07 & 90.6 $\pm$ 0.52 & 90.3 $\pm$ 6.77   \cr
        \hline
        
        \multirow{18}{*}{\rotatebox[origin=c]{90}{\textbf{Blackbox}}} & \multirow{3}{*}{Top2+True} & Precision & 49.8 $\pm$ 0.10 & 58.9 $\pm$ 1.45 & 62.4 $\pm$ 1.92 & 72.0 $\pm$ 3.56 & 62.3 $\pm$ 1.69 & 71.7 $\pm$ 1.95 & 69.4 $\pm$ 11.3 & 94.9 $\pm$ 3.55 \cr
        & & Recall & 59.6 $\pm$ 14.5 & 97.4 $\pm$ 2.53 & 95.3 $\pm$ 1.63 & 99.3 $\pm$ 0.74 & 71.1 $\pm$ 18.3 & \textcolor{blue1}{\textbf{99.9 $\pm$ 0.05}} &\textcolor{blue1}{\textbf{99.9 $\pm$ 0.08}}& 63.8 $\pm$ 35.9 \cr
        & & F1-Score & 52.1 $\pm$ 6.27 & 73.4 $\pm$ 0.41 & 75.4 $\pm$ 1.84 & 83.3 $\pm$ 2.24 & 62.9 $\pm$ 10.7 & 83.4 $\pm$ 1.29 & 80.9 $\pm$ 7.85 & 69.5 $\pm$ 25.6 \cr
        \cline{3-11}

        & \multirow{3}{*}{Loss-Threshold} & Precision & \textcolor{og}{\textbf{64.4 $\pm$ 2.07}} & 58.5 $\pm$ 0.72 & \textcolor{og}{\textbf{64.0 $\pm$ 1.45}} & 54.1 $\pm$ 26.6 & 61.6 $\pm$ 3.20 & \textcolor{og}{\textbf{79.4 $\pm$ 12.4}} & 76.0 $\pm$ 8.08 & \textcolor{og}{\textbf{98.4 $\pm$ 0.15}} \cr
        & & Recall & 49.9 $\pm$ 0.04 & 99.9 $\pm$ 0.02 & 82.6 $\pm$ 8.06 & 66.3 $\pm$ 2.03 & 52.8 $\pm$ 23.3 & 63.1 $\pm$ 8.94 & 99.2 $\pm$ 0.64 & 57.6 $\pm$ 17.1 \cr
        & & F1-Score & 56.2 $\pm$ 0.77 & 73.8 $\pm$ 0.57 & 71.8 $\pm$ 4.01 & 47.7 $\pm$ 19.7 & 48.1 $\pm$ 18.6 & 69.6 $\pm$ 9.60 & 85.6 $\pm$ 5.09 & 71.2 $\pm$ 13.7\cr
        \cline{3-11}
        
        & \multirow{3}{*}{Label-Only} & Precision &  46.5 $\pm$ 3.04 & 57.2 $\pm$ 0.10 & 55.1 $\pm$ 1.90 & 60.3 $\pm$ 0.16 & 56.4 $\pm$ 0.08 & 66.3 $\pm$ 0.70 & 75.3 $\pm$ 0.47 & 76.1 $\pm$ 1.26 \cr
        & & Recall & 71.7 $\pm$ 9.54 & \textcolor{blue1}{\textbf{99.9 $\pm$ 0.01}}& \textcolor{blue1}{\textbf{99.8 $\pm$ 0.10}} & 99.3 $\pm$ 0.47 &\textcolor{blue1}{\textbf{ 99.9 $\pm$ 0.04 }}& 99.9 $\pm$ 0.08 & 99.1 $\pm$ 0.47 & \textcolor{blue1}{\textbf{99.9 $\pm$ 0.01}} \cr
        & & F1-Score & 56.2 $\pm$ 5.28  & 72.8 $\pm$ 0.09 & 70.9 $\pm$ 1.54 & 75.3 $\pm$ 0.12 & 72.1 $\pm$ 0.07 & 79.7 $\pm$ 0.50 & 85.5 $\pm$ 0.47 & 86.4 $\pm$ 0.81\cr
        \cline{2-11}

        & \multirow{3}{*}{\textbf{BlindMI-Diff-w/}} & Precision & 50.0 $\pm$ 0.01 & \textcolor{og}{\textbf{70.8 $\pm$ 3.23}} & 64.0 $\pm$ 2.50 & \textcolor{og}{\textbf{78.1 $\pm$ 1.03}} & \textcolor{og}{\textbf{66.7 $\pm$ 0.78}} & 77.0 $\pm$ 0.53 & \textcolor{og}{\textbf{91.0 $\pm$ 0.17}} & 95.3 $\pm$ 0.49  \cr
        & & Recall &\textcolor{blue1}{\textbf{97.0 $\pm$ 1.18}} & 93.7 $\pm$ 0.47 & 97.9 $\pm$ 0.88 & 98.9 $\pm$ 0.93 & \textcolor{blue1}{\textbf{99.8 $\pm$ 0.07}}& 99.2 $\pm$ 0.54 & 98.9 $\pm$ 0.49 & 99.2 $\pm$ 0.59 \cr
        & & F1-Score & \textcolor{red}{\textbf{66.0} $\pm$ \textbf{0.28}} & \textcolor{red}{\textbf{80.6} $\pm$ \textbf{1.90}} & \textcolor{red}{\textbf{77.2} $\pm$ \textbf{1.83}} &\textcolor{red}{\textbf{87.3} $\pm$ \textbf{0.70}} & \textcolor{red}{\textbf{79.9} $\pm$ \textbf{0.57}} & \textcolor{red}{\textbf{86.7} $\pm$ \textbf{0.37}} & \textcolor{red}{\textbf{94.8} $\pm$ \textbf{0.14}} & \textcolor{red}{\textbf{97.2} $\pm$ \textbf{0.03}} \cr
        \cline{3-11}
        
        & \multirow{3}{*}{\textbf{BlindMI-Diff-w/o}} & Precision &50.0 $\pm$ 0.07 & 62.9 $\pm$ 0.10 & 62.3 $\pm$ 0.53 & 74.4 $\pm$ 0.61 & 63.6 $\pm$ 0.27 & 74.1 $\pm$ 1.62 & 88.2 $\pm$ 0.93 & 93.9 $\pm$ 0.50\cr 
        & & Recall & 90.1 $\pm$ 1.61& 99.6 $\pm$ 0.17 & 92.1 $\pm$ 1.29 & \textcolor{blue1}{\textbf{99.9 $\pm$ 0.01}}& 99.5 $\pm$ 0.02& 97.4 $\pm$ 1.83 & 98.8 $\pm$ 0.68 & 99.0 $\pm$ 0.76 \cr
        & & F1-Score & 64.2 $\pm$ 0.27 & 77.1 $\pm$ 0.13 & 74.3 $\pm$ 0.80 & 85.3 $\pm$ 0.39 & 77.6 $\pm$ 0.19 & 84.1 $\pm$ 0.42 & 93.2 $\pm$ 0.82 & 96.4 $\pm$ 0.09 \cr
        \cline{3-11}
        
        & \multirow{3}{*}{\textbf{BlindMI-1Class}} & Precision & 50.0 $\pm$ 0.08 & 67.3 $\pm$ 2.38 & 62.5 $\pm$ 2.80 & 67.3 $\pm$ 1.66 & 64.4 $\pm$ 0.39 & 71.5 $\pm$ 1.57 & 90.8 $\pm$ 0.68 & 95.8 $\pm$ 0.53 \cr
        & & Recall & 69.9 $\pm$ 0.04 & 96.6 $\pm$ 0.62 & 96.2 $\pm$ 1.02 & 99.9 $\pm$ 0.03 & 99.0 $\pm$ 0.22 & 99.3 $\pm$ 0.06 & 97.6 $\pm$ 0.39 & 97.1 $\pm$ 1.88 \cr
        & & F1-Score & 58.3 $\pm$ 0.07 & 79.3 $\pm$ 1.44 & 75.6 $\pm$ 1.90 & 80.4 $\pm$ 1.19 & 78.0 $\pm$ 0.32 & 83.4 $\pm$ 1.03 & 94.0 $\pm$ 0.47  & 96.4 $\pm$ 0.33  \cr
        \hline

         \multirow{18}{*}{\rotatebox[origin=c]{90}{\textbf{Gray-Blind}}} & \multirow{3}{*}{NN} & Precision & 50.2 $\pm$ 0.14 & 62.1 $\pm$ 0.82 & 62.2 $\pm$ 0.87 & 74.8 $\pm$ 1.14 & 64.2 $\pm$ 0.57 & 72.4 $\pm$ 1.75 & 90.4 $\pm$ 1.02 & 95.7 $\pm$ 0.56 \cr
        & & Recall & 64.1 $\pm$ 16.7 & 96.1 $\pm$ 2.23 & 90.2 $\pm$ 4.21 & 99.8 $\pm$ 0.04 & 96.4 $\pm$ 1.65 & 98.5 $\pm$ 1.22 & 96.2 $\pm$ 0.27 & 97.8 $\pm$ 0.05 \cr
        & & F1-Score & 54.3 $\pm$ 5.50 & 72.3 $\pm$ 0.08 & 73.5 $\pm$ 1.99 & 85.6 $\pm$ 0.71 & 77.0 $\pm$ 0.36 & 83.4 $\pm$ 0.83 & 93.2 $\pm$ 0.46 & 96.8 $\pm$ 0.28\cr 
        \cline{3-11}
    
        & \multirow{3}{*}{Top3-NN} & Precision & 50.2 $\pm$ 0.21 & 60.7 $\pm$ 1.10 & 62.0 $\pm$ 0.77 & 75.0 $\pm$ 1.06 & 64.1 $\pm$ 0.47 & 72.0 $\pm$ 1.65 & 90.4 $\pm$ 0.86 & 89.9 $\pm$ 0.08   \cr
        & & Recall & 69.9 $\pm$ 25.7 & 97.5 $\pm$ 1.56 & 90.7 $\pm$ 3.87 & 99.9 $\pm$ 0.05 & 97.2 $\pm$ 1.14 & \textcolor{blue1}{\textbf{99.1 $\pm$ 0.69}} & 96.3 $\pm$ 0.15& 96.8 $\pm$ 0.14\cr
        & & F1-Score & 56.4 $\pm$ 9.27 & 74.8 $\pm$ 0.37 & 73.6 $\pm$ 1.80 & 85.7 $\pm$ 0.69 & 77.2 $\pm$ 0.34 & 83.4 $\pm$ 0.90 & 93.2 $\pm$ 0.80 & 93.2 $\pm$ 0.03 \cr
        \cline{3-11}
        
        & \multirow{3}{*}{Top1-Threshold} & Precision & 0.51 $\pm$ 0.23 & \textcolor{og}{\textbf{99.9 $\pm$ 0.03}}& \textcolor{og}{\textbf{74.7 $\pm$ 24.9}} & 14.1 $\pm$ 2.93 & 61.6 $\pm$ 12.1  & 0.34 $\pm$ 0.20 & \textcolor{og}{\textbf{97.2 $\pm$ 1.70}} & \textcolor{og}{\textbf{99.9 $\pm$ 0.06}} \cr
        & & Recall & 31.3 $\pm$ 13.6 & 55.2 $\pm$ 0.51 & 40.9 $\pm$ 13.7 & 63.7 $\pm$ 1.60 & 63.2 $\pm$ 1.97 & 24.6 $\pm$ 14.2 & 86.2 $\pm$ 3.55 & 57.9 $\pm$ 0.81\cr
        & & F1-Score & 1.01 $\pm$ 0.44 & 71.1 $\pm $ 0.42 & 52.8 $\pm$ 17.6 & 22.7 $\pm$ 3.87 & 53.5 $\pm$ 7.26 & 0.67 $\pm$ 0.38 & 92.8 $\pm$ 1.72 & 71.4 $\pm$ 0.65 \cr
        \cline{2-11}
        
        & \multirow{3}{*}{\textbf{BlindMI-Diff-w/}} & Precision & 50.0 $\pm$ 0.03 & 66.2 $\pm$ 2.80 & 65.4 $\pm$ 2.52 & \textcolor{og}{\textbf{75.9 $\pm$ 1.44}} & \textcolor{og}{\textbf{66.0 $\pm$ 0.55}} & \textcolor{og}{\textbf{76.1 $\pm$ 1.06}} & 90.7 $\pm$ 0.86 & 95.5 $\pm$ 0.08 \cr
        & & Recall & \textcolor{blue1}{\textbf{90.4 $\pm$ 5.95}}& 94.4 $\pm$ 3.35 & 88.8 $\pm$ 2.04 & 99.9 $\pm$ 0.08 & 95.2 $\pm$ 1.38 & 97.4 $\pm$ 1.64 & 97.4 $\pm$ 0.37 & 98.2 $\pm$ 0.26\cr
        & & F1-Score & \textcolor{red}{\textbf{64.2 $\pm$ 1.59}}  & \textcolor{red}{\textbf{77.7} $\pm$ \textbf{0.80}}  & \textcolor{red}{\textbf{75.1} $\pm$ \textbf{1.49}} &  \textcolor{red}{\textbf{86.2} $\pm$ \textbf{0.90}} & \textcolor{red}{\textbf{78.0} $\pm$ \textbf{0.31}} & \textcolor{red}{\textbf{85.5 $\pm$ 0.80}} & \textcolor{red}{\textbf{93.9} $\pm$ \textbf{0.63}} & \textcolor{red}{\textbf{96.8} $\pm$ \textbf{0.09}} \cr
        \cline{3-11}
        
        & \multirow{3}{*}{\textbf{BlindMI-Diff-w/o}} & Precision & \textcolor{og}{\textbf{50.5 $\pm$ 0.05}} & 60.3 $\pm$ 1.26 & 63.3 $\pm$ 1.60 & 71.5 $\pm$ 0.87 & 61.7 $\pm$ 0.38 & 69.2 $\pm$ 4.09 & 86.1 $\pm$ 0.78 & 94.5 $\pm$ 0.24 \cr 
        & & Recall & 84.2 $\pm$ 4.05 & \textcolor{blue1}{\textbf{99.3 $\pm$ 0.10}} &  92.4 $\pm$ 2.68 & \textcolor{blue1}{\textbf{99.9 $\pm$ 0.04}} & \textcolor{blue1}{\textbf{99.5 $\pm$ 0.25}}& 97.6 $\pm$ 1.83& \textcolor{blue1}{\textbf{98.9 $\pm$ 0.60}} & 98.0 $\pm$ 0.80 \cr
        & & F1-Score & 62.7 $\pm$ 1.12 & 75.0 $\pm$ 1.40 & 75.1 $\pm$ 1.89 & 83.3 $\pm$ 0.57 & 76.2 $\pm$ 0.25 & 80.7 $\pm$ 2.37 & 92.1 $\pm$ 1.15 & 96.2 $\pm$ 0.26 \cr
        \cline{3-11}
        
        & \multirow{3}{*}{\textbf{BlindMI-1Class}} & Precision & 49.9 $\pm$ 0.07 & 64.7 $\pm$ 2.41 & 59.0 $\pm$ 1.13 & 63.4 $\pm$ 0.44 & 61.1 $\pm$ 1.38 & 70.9 $\pm$ 1.11 & 83.5 $\pm$ 0.40 & 83.8 $\pm$ 11.9 \cr
        & & Recall & 56.5 $\pm$ 5.85 & 94.4 $\pm$ 0.01 & \textcolor{blue1}{\textbf{94.8 $\pm$ 1.41}} & 99.7 $\pm$ 0.27 & 98.1 $\pm$ 0.71 & 97.2 $\pm$ 2.07 & 98.9 $\pm$ 0.69 & \textcolor{blue1}{\textbf{99.1 $\pm$ 0.60}} \cr
        & & F1-Score & 52.6 $\pm$ 2.47 & 76.8 $\pm$ 1.70 & 72.8 $\pm$ 1.27 & 77.6 $\pm$ 0.33 & 77.1 $\pm$ 0.50 & 81.9 $\pm$ 0.07 & 90.6 $\pm$ 0.52 & 90.3 $\pm$ 6.77   \cr
        \hline

        \multirow{18}{*}{\rotatebox[origin=c]{90}{\textbf{Graybox}}} & \multirow{3}{*}{Top2+True} & Precision &  50.0 $\pm$ 0.04 &
        63.7 $\pm$ 1.53 &
        62.8 $\pm$ 1.65 &
        75.4 $\pm$ 0.85 &
        65.1 $\pm$ 0.38 &
        75.1 $\pm$ 0.28 &
        90.6 $\pm$ 0.68 &
        95.7 $\pm$ 0.48\cr
        & & Recall &  96.8 $\pm$ 2.22 &
        98.4 $\pm$ 1.38 &
        93.4 $\pm$ 2.95 &
        99.9 $\pm$ 0.04 &
        98.3 $\pm$ 0.66 &
        \textcolor{blue1}{\textbf{99.9 $\pm$ 0.01}} &
        97.3 $\pm$ 0.42 &
        98.1 $\pm$0.13
        \cr
        & & F1-Score & 66.0 $\pm$ 0.50 & 77.3 $\pm$ 0.69 & 75.1 $\pm$ 2.03 & 86.0 $\pm$ 0.55 & 78.4 $\pm$ 0.25 & 85.7 $\pm$ 0.18 & 93.8 $\pm $0.53 & 96.9 $\pm$ 0.18 \cr
        \cline{3-11}

        & \multirow{3}{*}{Loss-Threshold} & Precision &  \textcolor{og}{\textbf{66.4 $\pm$ 2.38}} &
        \textcolor{og}{\textbf{96.0 $\pm$ 3.93}} &
        \textcolor{og}{\textbf{84.4 $\pm$ 4.83}} &
        77.2 $\pm$ 10.5 &
        \textcolor{og}{\textbf{79.1 $\pm$ 5.41}} &
        75.4 $\pm$ 9.79 &
        84.9 $\pm$ 8.05 &
        75.4 $\pm$ 1.26\cr
        & & Recall & 50.0 $\pm$ 0.07 &
        64.1 $\pm$ 2.70 &
        64.5 $\pm$ 2.07 &
        75.6 $\pm$ 1.79 &
        66.1 $\pm$ 0.85 &
        79.5 $\pm$ 0.64 &
        90.8 $\pm$ 3.06 &
        98.2 $\pm$ 0.23\cr
        & & F1-Score & 57.0 $\pm$ 0.84 & 76.8 $\pm$ 0.68 & 73.0 $\pm$ 2.90 & 75.9 $\pm$ 4.96 & 71.8 $\pm$ 2.70 & 76.5 $\pm$ 4.81 & 87.1 $\pm$ 3.39 & 85.3 $\pm$ 0.89\cr
        \cline{3-11}
        
        & \multirow{3}{*}{Label-Only} & Precision & 46.5 $\pm$ 0.30 &
        57.2 $\pm$ 0.10 &
        55.1 $\pm$ 1.90 &
        60.4 $\pm$ 0.16 &
        56.4 $\pm$ 0.08 &
        66.3 $\pm$ 0.90 &
        75.3 $\pm$ 0.47 &
        76.1 $\pm$1.26\cr
        & & Recall & 71.7 $\pm$ 9.54 &
        \textcolor{blue1}{\textbf{99.9 $\pm$ 0.01}} &
        \textcolor{blue1}{\textbf{99.8 $\pm$ 0.10}} &
        99.3 $\pm$ 0.47 &
        \textcolor{blue1}{\textbf{99.9 $\pm$ 0.04}} &
        99.9 $\pm$ 0.08 &
        \textcolor{blue1}{\textbf{99.1 $\pm$ 0.47}} &
        57.9 $\pm$ 0.81\cr
        & & F1-Score & 56.2 $\pm$ 5.28 & 
        72.8 $\pm$ 0.09 & 
        70.9 $\pm$ 1.54 &
        75.3 $\pm$ 0.12 &
        72.1 $\pm$ 0.07 & 
        79.7 $\pm$ 0.50 & 
        85.5 $\pm$ 0.47 & 
        86.4 $\pm$ 0.81\cr
        \cline{2-11}

        & \multirow{3}{*}{\textbf{BlindMI-Diff-w/}} & Precision & 50.0 $\pm$ 0.01 &
        70.8 $\pm$ 3.23 &
        64.0 $\pm$ 2.50 &
        \textcolor{og}{\textbf{78.1 $\pm$ 1.0}} &
        66.7 $\pm$ 0.78 &
        \textcolor{og}{\textbf{77.0 $\pm$ 0.53}} &
        \textcolor{og}{\textbf{91.1 $\pm$ 0.17}} &
        95.3 $\pm$ 0.49\cr
        & & Recall & \textcolor{blue1}{\textbf{97.0 $\pm$ 1.18}}&
        93.7 $\pm$ 0.47 &
        97.7 $\pm$ 0.88 &
        98.9 $\pm$ 0.93 &
        99.8 $\pm$ 0.07 &
        99.2 $\pm$ 0.54 &
        98.9 $\pm$ 0.49 &
        \textcolor{blue1}{\textbf{98.2 $\pm$ 0.26}}\cr
        & & F1-Score & \textcolor{red}{\textbf{66.0} $\pm$ \textbf{0.30}} & 
        \textcolor{red}{\textbf{80.6} $\pm$ \textbf{1.90}} & \textcolor{red}{\textbf{77.2} $\pm$ \textbf{1.83}} & \textcolor{red}{\textbf{87.3} $\pm$ \textbf{0.70}} &   \textcolor{red}{\textbf{79.9} $\pm$ \textbf{0.57}} & 
        \textcolor{red}{\textbf{86.7} $\pm$ \textbf{0.37}} & 
        \textcolor{red}{\textbf{94.8} $\pm$ \textbf{0.14}} & \textcolor{red}{\textbf{97.2} $\pm$ \textbf{0.03}} \cr
        \cline{3-11}
        
        & \multirow{3}{*}{\textbf{BlindMI-Diff-w/o}} & Precision &50.0 $\pm$ 0.07 & 62.9 $\pm$ 0.10 & 62.3 $\pm$ 0.53 & 74.4 $\pm$ 0.61 & 63.6 $\pm$ 0.27 & 74.1 $\pm$ 1.62 & 88.2 $\pm$ 0.93 & 93.9 $\pm$ 0.50\cr 
        & & Recall & 90.1 $\pm$ 1.61& 99.6 $\pm$ 0.17 & 92.1 $\pm$ 1.29 & \textcolor{blue1}{\textbf{99.9 $\pm$ 0.01}}& 99.5 $\pm$ 0.02& 97.4 $\pm$ 1.83 & 98.8 $\pm$ 0.68 & 99.0 $\pm$ 0.76 \cr
        & & F1-Score & 64.2 $\pm$ 0.27 & 77.1 $\pm$ 0.13 & 74.3 $\pm$ 0.80 & 85.3 $\pm$ 0.39 & 77.6 $\pm$ 0.19 & 84.1 $\pm$ 0.42 & 93.2 $\pm$ 0.82 & 96.4 $\pm$ 0.09 \cr
        \cline{3-11}
        
        & \multirow{3}{*}{\textbf{BlindMI-1Class}} & Precision & 50.0 $\pm$ 0.08 &
        67.3 $\pm$ 2.38 &
        62.5 $\pm$ 2.80 &
        67.3 $\pm$ 1.66 &
        64.4 $\pm$ 0.39 &
        71.5 $\pm$ 1.57 &
        90.8 $\pm$ 0.68 &
        \textcolor{og}{\textbf{95.8 $\pm$ 0.53}}\cr
        & & Recall & 69.9 $\pm$ 0.04 &
        96.6 $\pm$ 0.62 &
        96.2 $\pm$ 1.02 &
        99.9 $\pm$ 0.03&
        99.0 $\pm$ 0.22 &
        99.9 $\pm$ 0.06 &
        97.6 $\pm$ 0.39 &
        97.4 $\pm$ 0.05\cr
        & & F1-Score & 58.3 $\pm$ 0.07 &
        79.3 $\pm$ 1.44 &
        75.6 $\pm$ 1.90 &
        80.4 $\pm$ 1.19 &
        78.0 $\pm$ 0.32 &
        83.4 $\pm$ 1.03 &
        94.0 $\pm$ 0.47 &
        96.4 $\pm$ 0.33 \cr

        \bottomrule
	\end{tabular} \vspace{-0.15in}
\end{table*}

Table~\ref{table_results} shows the Precision, Recall and F1-score of different attacks under four adversarial settings.
\textcolor{blue}{The best performances of all attacks under different settings are highlighted with different colors (blue for recall, green for precision, and red for F1-score.)}
Note that if the performance of a prior attack, e.g., NN-based, is the same with and without ground truth label, we only show the attack once under the blind and graybox-blind settings.  We do show \sys multiple times under different settings for ease of comparison.  Next, we introduce several observations from our experiments.


[Observation RQ1-1] \textit{\sys significantly outperforms state-of-the-art MI attacks under all settings in terms of F1-score.}

The first observation is that \sys outperforms state-of-the-art MI attacks under all settings: The reason is that \sys extracts membership semantics directly from the target model via probing.  Sometimes, the performance boost is over 20\%, e.g., for \highlight{the Adult} and BIRDS-200 \highlight{datasets} under the blind setting. 
 As a comparison, no single prior attack dominates the performance in F1-score. Consider the blind setting for example. Top1-Thre is the best for the EyePACS dataset except for \sys; NN is the best for the CH-MNIST dataset except for \sys; and Top3-NN outperforms all methods for the Purchase-50.  The reason is that no prior attacks extract enough membership semantics as \sys does. 

[Observation RQ1-2] \textit{The introduction of ground-truth labels improves attack performance, but to a limited degree for \sys.}

The second observation is about how the introduction of ground-truth labels affects attack performance. The performance boost is sometimes significant for prior attacks. Take Purchase-50 for example.  The best average F1-score under the blind setting is 59.6\%, but the average F1-score increases to 72.1\%, a 12.5\% increase, under the blackbox setting. 

As a comparison, the best performance boost of \sys with the ground truth label is 2.9\% for the EyePACS dataset.  That said, although ground-truth labels introduce additional membership semantics, the semantics introduction is limited in terms of F1-score improvement. 

[Observation RQ1-3] \textit{Shadow model quality plays an important role in some existing attacks.}

The third observation is about how different shadow models affect the attack performance.  First, \sys does not need a shadow model and therefore \sys's performance is the same with or without shadow model.  Second, the performance of some existing attacks varies a lot given different shadow models. Take the NN attack for BIRDS-200 under the blind setting for example.  The average F1-score is 58.3, but the standard error is 15.0 with a confidence of 68.3\%.  That said, the choice of shadow models is crucial in the performance of existing attacks with shadow models.

[Observation RQ1-4] \textit{\sysdiff-w/ performs the best among all three variations in terms of F1-score, while \sysdiff-w/o does not need additional probes to the target model.}

Table~\ref{table_results} shows that \sysdiff-w/ is the best comparing with \sysdiff-w/o and \syssvm. At the same time, \sysdiff-w/ does require 20 additional probes to the target model to generate a non-member set. If the adversary's access to the target model is restricted to the target dataset,  \sysdiff-w/o is an alternative option as opposed to \sysdiff-w/. 

[Observation RQ1-5] \textit{The variation of \syssvm is larger than the one of \sysdiff.}

Take Birds-200 for example. The best performance of \syssvm is higher than the one of \sysdiff. The reason is that the performance of ML classifier depends on the training data: If many data samples lie along the decision boundary, the one-class model can learn the membership semantics and thus outperforms \sysdiff.

\subsection{RQ2: Defenses} \label{RQ2}


In this subsection, we evaluate the performance of \sys against state-of-the-art defenses of MI attacks.  Note that we evaluate all the defenses under the blackbox setting because some attacks only work under the blackbox but not the blind setting. Now, we describe three general defense directions and  representative works in each direction below. 


\begin{icompact}
\item Output probability alteration based on adversarial example.  \hspace{0.05in} Such a defense alters the output probabilities so that it becomes hard for an adversary to infer membership information. A representative approach in this category is called MemGuard~\cite{memguard}, which changes the output probability distribution so that it looks like an adversarial example to the inference model built by the adversary.  We adopt the original implementation of MemGuard.\footnote{\url{https://github.com/jjy1994/MemGuard}}
\item Regularization-based fortification of ML model. \hspace{0.05in} Such a defense fortifies existing ML models, especially DNN, via regularization.  Two representative approaches in this category are MMD+Mix-up~\cite{mixup} (which include two previous defenses, namely dropout~\cite{globalloss} and L2-Regularizer~\cite{NDSS}) and the adversarial regularization~\cite{minmax}.  We implement our own version of MMD+Mix-up and adopt an open-source version of the adversarial regularization.\footnote{\url{https://github.com/NNToan-apcs/python-DP-DL}}  \highlight{Note that the MMD+Mixup defense is adaptive with a regularizer based on BlindMI, i.e., minimizing the cluster distance between members and non-members during the training process. As a comparison,  the adversarial regularization is based on the NN attack, because it requires that the MI attack be differentiable with gradients while BlindMI is not.  (If we adopt a differentiable distance function in adversarial regularization, adversarial regularization boils down to the MMD+Mixup method.)}

\item Differential privacy-based protection. \hspace{0.05in} Such a defense adds noise to the output to fool an adversary.  A representative approach in this category is DP-Adam~\cite{DP} and we adopt an open-source version of DP-Adam.\footnote{\url{https://github.com/tensorflow/privacy}}
\end{icompact}

Note that in our experiment, we adopt a dataset that is at least included in the corresponding defense paper for our evaluation. That is, we choose CH-MNIST for MemGuard and DP-Adam, and CIFAR-100 for MMD+Mix-up and Adversarial Regularization. Because these defenses adopted different datasets in the paper and we follow what what adopted.


\subsubsection{Attacks against MemGuard}
We first evaluate the performance of MemGuard under existing MI attacks. The utility-loss budget is set up as [0, 0.1, 0.3, 0.5, 0.7, 1.0] for MemGuard, which represents the percentage of altered outputs. We show the evaluation results in Figure~\ref{fig_memguard} and also make the following observations.

[Observation RQ2-1] \textit{Attacks with ground-truth labels generally have a higher F1-score than those without when attacking MemGuard.}

Our first observation is that MemGuard is generally vulnerable to attacks that utilize ground-truth labels.  For example, the worst performing attacks are Top1-Threshold and Top3-NN, which will likely remove the output probability of the ground-truth labels. By contrast, the best performed attacks are Label-only, Top2+True, and \sys, which are all able to utilize the ground-truth label information. The reason is that although MemGuard alters the output probabilities, it does not change the prediction class because MemGuard does not want to influence the legacy performance of the model. 

There are two more things worth noting. First, the performance of NN is actually better than the one of Top3-NN.  The reason is similarly: NN adopts all the probability scores of the output, which contains the one corresponding to the ground-truth label for certain; by contrast, Top3-NN only adopts the top three probability scores, which may not contains the one corresponding to the ground-truth label.  

Second, Top2+True performs better than Label-only when the privacy budget is small, but then degrades quickly when the privacy budget increases. The reason is that when the budget is small, top two probability scores will provide some membership information.  When the budget increases, the top two of more samples are altered, which affects the performance of Top2+True.  


[Observation RQ2-2] \textit{\sys still outperforms all existing attacks even if the output probabilities were adversarially altered.}

Our second observation is that \sys still performs the best among all attacks. The underlying reasons are two-fold.  First, although adversarial examples are close to the decision boundary, the decision boundary itself is a hyper-dimensional manifold and the projection of members and non-members on the manifold are still far from each other, thus being distinguishable. Second, although MemGuard alters output probability scores, sufficient information still exist, because MemGuard does not change the prediction results. 

\subsubsection{Attacks against DP-Adam}
In this part, we evaluate all existing attacks against DP-Adam under the 
 noise\_multiplier as [0, 0.002, 0.004, 0.006, 0.008, 0.01].  The evaluation results are shown in Figure~\ref{fig_dp} and we make the following observations.

[Observation RQ2-3] \textit{Attacks relying on binary comparison tend to have a low F1-score against DP-Adam.}

The reason behind this observation is that differential privacy (DP) perpetuates the probability outputs so that the boundary between members and non-members is blurred. Therefore, the performances of Top1-Threshold and Loss-Threshold are the worst.  Consider Top1-Threshold for example: It is hard to differentiate members and non-members based on a single threshold of the highest output probability score due to the perpetuation enforced by DP. 


[Observation RQ2-4] \textit{\sys has a higher performance than Label-only regardless of when the privacy-utility budget is small or large, while Label-only attack degrades slower with a mid-size budget. }


The reason is that Label-only attack depends on the performance gap of the target model on the training and testing datasets.  Such a gap persists with a mid-size budget, but starts to shrink quickly for a large budget---and that is why Label-only's performance degrades finally with a large budget.   

%

\subsubsection{Attacks against MMD+Mixup} In this part, we evaluate all the MI attacks against MMD+Mixup with different privacy-utility budgets, i.e., the loss weight in the MMD as [0, 0.1, 0.5, 1, 2.5, 5]. 
 Note that this budget controls the tradeoff between privacy and utility: A larger privacy-utility budget increases privacy protection, but at the same time decreases the model's utility. 
%
%
 The evaluation results are in Figure~\ref{fig_mmdmixup}---\sys clearly outperforms all existing attacks.  We also make the following observation.


[Observation RQ2-5] \textit{Attacks selecting more probability scores generally have a higher F1-score than those selecting less when attacking MMD+Mix-up.}

Specifically, Label-Only and Top1-Threshold are the worst when comparing with other attacks.  The reason is that MMD+Mix-up also changes the target model's performance on the training dataset and therefore Label-Only and Top1-Threshold are heavily affected.  As a comparison, other attacks also rely on the probability score of other classes, thus outperforming these two attacks. 

\subsubsection{Attacks against Adversarial Regularization} 
\begin{figure*}[t!]
\centering

\subfigure[MemGuard on CH-MNIST]{
\begin{minipage}[t]{0.5\linewidth}
\centering
\includegraphics[width=0.93\linewidth]{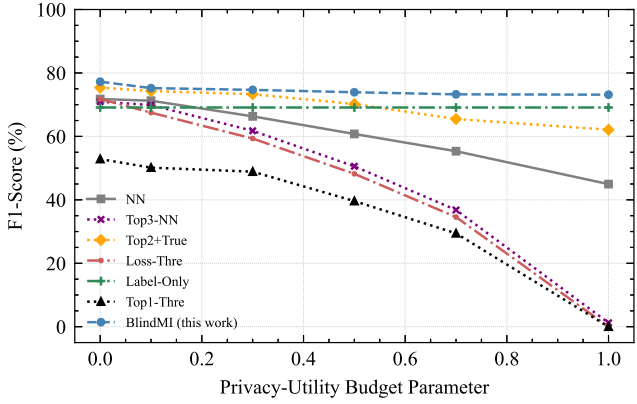}
\graphicspath{{../eps/}}
\DeclareGraphicsExtensions{.eps}
\label{fig_memguard}
\end{minipage}%
}%
\subfigure[DP-Adam on CH-MNIST]{
\begin{minipage}[t]{0.5\linewidth}
\centering
\includegraphics[width=0.93\linewidth]{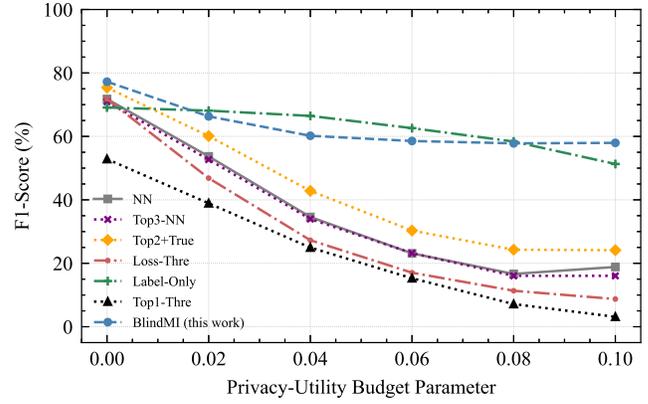}
\graphicspath{{../eps/}}
\DeclareGraphicsExtensions{.eps}
\label{fig_dp}
\end{minipage}%
}%
\quad
\subfigure[MMD+Mix-up on CIFAR-100]{
\begin{minipage}[t]{0.5\linewidth}
\centering
\includegraphics[width=0.93\linewidth]{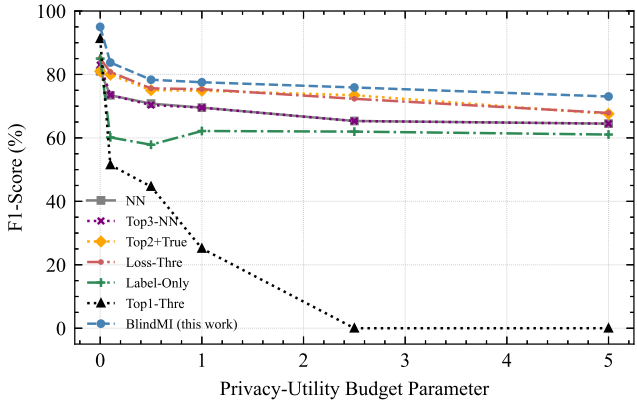}
\graphicspath{{../eps/}}
\DeclareGraphicsExtensions{.eps}
\label{fig_mmdmixup}
\end{minipage}
}%
\subfigure[Adversarial Regularization on CIFAR-100]{
\begin{minipage}[t]{0.5\linewidth}
\centering
\includegraphics[width=0.93\linewidth]{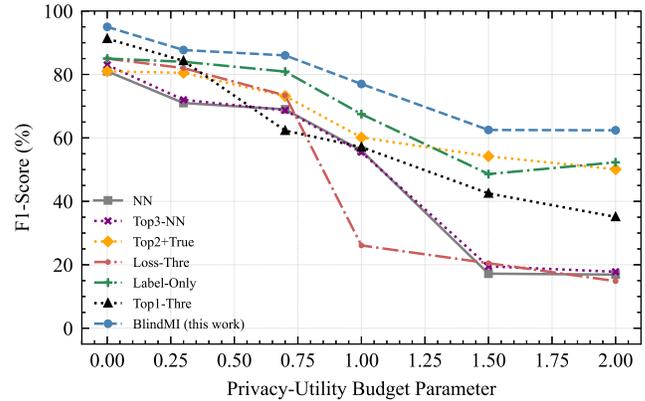}
\graphicspath{{../eps/}}
\DeclareGraphicsExtensions{.eps}
\label{fig_ar}
\end{minipage}
}%
\centering \vspace{-0.1in}
\caption{F1-Score of Different MI attacks, i.e., state-of-the-art and \sys, against a Target Model together with Corresponding Defenses.} 
\label{against_defense}
\end{figure*}

In this part, we evaluate all the MI attacks against the Adversarial Regularization~\cite{minmax} with different privacy-utility budget as [0, 0.3, 0.7, 1, 1.5, 2].  \sys clearly outperforms all existing attacks as shown in Figure~\ref{fig_ar}.  Now we describe our observations. 

[Observation RQ2-6] \textit{Ground-truth Label plays an important role in defeating Adversarial Regularization and the results depend on how such labels are used in the attack.}

Specifically, Label-Only, Top2+True and \sys, which all adopt ground-truth labels, are the best three MI attacks among all, while Loss-Threshold, which also adopts ground-truth labels, is the worst.  The reason is that Loss-Threshold relies on the training data of a shadow model, which is drastically different from a \highlight{model} trained with adversarial regularization. 

[Observation RQ2-7] \textit{Simple MI attacks, except for \sys, tend to have a better performance.}

Specifically, Label-Only and Top1-Threshold performs well compared with other attacks.  The reason is that although the adversarial regularization fortifies the model via regularization, the output probability, especially the probability score of the predicted class, still contains abundant information.

\subsection{RQ3: Nonmember Set Quality and Size} \label{sub:size}

In this subsection, we evaluate how different generation methods and size of non-members affect the performance of \sys, particularly \sysdiff and \syssvm. Without loss of generality, we use the EyePACS dataset and the blind setting: \highlight{the target dataset consists of 20,000 samples} and the size of non-member datasets changes from 20 to 10,000. Here are the settings used in the different non-member generation methods:

\begin{icompact}
\item Sample transformation. \hspace{0.01in} We adopt the Sobel operator. 
\item Random perpetuation. \hspace{0.01in} We adopt Gaussian noise with the mean value as zero and the variance as 0.001. 
\item Random generation. \hspace{0.01in} We adopt a uniform distribution in generating feature values. 
\item Cross-domain sample. \hspace{0.01in} We adopt samples from CH-MNIST. 
\end{icompact}

We show our experiment results in Table~\ref{tab:setqualityandsize} and also make the following observations. 




\begin{table*}[!t]
   \renewcommand{\arraystretch}{1.2}
   \setlength{\tabcolsep}{3pt}
   \scriptsize
    \caption{F1-Score(\%) of \sysdiff/\syssvm with standard error of the mean of with nonmember datasets generated via different methods of varied size. } \vspace{-0.05in}
    \label{tab:setqualityandsize}
	\centering
	\begin{tabular}{lcccccc}
	    \toprule 
	    Method $\backslash$ Size & 20 & 50 & 100 & 200 & 1,000 & 10,000 \\
	    	    \midrule
	    Sample transform & 77.7$\pm$0.80 / 72.9$\pm$1.82 & 78.1$\pm$0.99 / 74.4$\pm$1.43 & 77.9$\pm$1.58 / 75.6$\pm$1.55 & 78.2$\pm$0.87 / 76.3$\pm$1.67 & 78.4$\pm$1.01 / 76.8$\pm$1.70 & \textbf{78.7$\pm$0.63} / \textbf{77.5$\pm$1.31}\\
	    Random perpetuation & 77.5$\pm$1.37 / 66.4$\pm$1.20 & 77.9$\pm$0.92 / 72.1$\pm$1.30 & 77.6$\pm$1.44 / 73.1$\pm$1.07 & 78.0$\pm$0.85 / 73.5$\pm$0.70 & 77.6$\pm$0.42 / 74.5$\pm$0.88 & \textbf{78.2$\pm$0.30} / \textbf{75.7$\pm$0.69}\\
	    Random generation & 75.5$\pm$2.51 / 71.6$\pm$1.98 & 75.7$\pm$1.93 / 71.6$\pm$2.31 & 75.3$\pm$2.03 / 71.3$\pm$2.38 & 75.6$\pm$1.79 / 71.4$\pm$2.00 & 75.7$\pm$1.59 / 71.8$\pm$2.03 & \textbf{75.7$\pm$1.64} / \textbf{72.2$\pm$1.87}\\
	    Cross domain & 77.9$\pm$1.26 / 64.9$\pm$1.99 & 78.0$\pm$1.38 / 71.4$\pm$1.46 & \textbf{78.1$\pm$1.21 }/ 72.5$\pm$1.60 & 78.1$\pm$1.05 / 73.2$\pm$1.09 & 77.8$\pm$1.20 / 76.1$\pm$1.33 & 77.6$\pm$1.37 / \textbf{77.0$\pm$0.93}\\
	    \bottomrule
	 \end{tabular} 
\end{table*}

%
%
%

\begin{table*}[!t]
   \renewcommand{\arraystretch}{1.2}
   \scriptsize
    \caption{\textcolor{blue}{Mann-Whitney U test value (P-value) of F-1 scores of \sysdiff/\syssvm with nonmember sets via different methods}} 
   \vspace{-0.05in}
    \label{BlindMi_Pvalue}
	\centering
		\begin{threeparttable}
	\begin{tabular}{ccccc}
	    \toprule 
	    & Sample transform & Random perpetuation & Random generation & Cross domain \cr
	    \midrule
	    Sample transform & 18$^*$ (0.4678$^{**}$) / 18 (0.4678)& -- & -- & -- \cr
	    Random perpetuation & 7 (0.0455) / 7 (0.0463) & 18 (0.4673) / 18 (0.4678) & -- & -- \cr
	    Random generation & 0 (0.0024) / 0 (0.0025) & 0 (0.0023) / 7 (0.0461) & 18 (0.4657) / 18 (0.4673) & -- \cr
	    Cross domain &  9.5 (0.0981) / 7 (0.0463) & 13 (0.2328) / 18 (0.4681) & 0 (0.0023) / 10.5 (0.1303) & 18 (0.4673) / 18 (0.4678)\cr
    \bottomrule
	 \end{tabular} 
	 \begin{tablenotes}
        \footnotesize
        \item[*]: the larger the U value is, the more similar two datasets are.  
        \item[**]: a p-value less than 0.05 indicates statistical significance.
      \end{tablenotes} 
\end{threeparttable}
\end{table*}

\begin{table}[!t]
   \renewcommand{\arraystretch}{1.2}
   \setlength{\tabcolsep}{2.5pt}
   \scriptsize
    \caption{\textcolor{blue}{MMD statistical tests of \sysdiff with nonmember datasets generated via different methods (each value is the MMD with standard error of the mean between corresponding samples and real-world non-members in the test dataset.) 
     }}
    \label{MMD_distance}
	\centering
	\begin{tabular}{c|c|c|c|c}
	    \toprule 
	     Sample trans & Random perp &  Random generation &Cross domain &Training set \cr
	    \midrule
	    \textbf{0.194 $\pm$ 0.009} & 0.438 $\pm$ 0.039 & 3.024 $\pm$ 1.024& 0.225 $\pm$ 0.015 &1.864 $\pm$ 0.022  \cr
	    
	    \bottomrule
	 \end{tabular} \vspace{-0.15in}
\end{table}

[Observation RQ3-1] \textit{The performance of \sysdiff stays mostly stable with a little increase as opposed to a big increase of \syssvm as the size of nonmember datasets increases.}

Our first observation of RQ3 is on how the size of nonmembers affects F1-score. The F1-score of \sysdiff is almost constant with around 1\% boost as the size increases from 20 to 10,000.  As a comparison, the F1-score of \syssvm has between 5\% and 12\% increase except for random generated non-members.  

The reason is that \syssvm adopts a learning model, particularly one-class SVM, which needs some training data to learn the underlying semantics.  As a comparison, \sysdiff directly compares the distribution between the target and the non-member, which is effective in extracting membership semantics from just a few samples. 

[Observation RQ3-2] \textit{The quality of sample transformation is the best, while the random generation of non-members is the worst among all four methods.}


\textcolor{blue}{Our second observation is on how different nonmember generation methods affect attack performance.  Table~\ref{tab:setqualityandsize} shows that sample transformation is the most effective method for both \sysdiff and \syssvm. Since the differences are relatively small, we perform two statistical tests, i.e., (i) the Mann-Whitney U test and the P-value, and (ii) the maximum mean discrepancy (MMD) tests, to demonstrate the statistical significance.}

\highlight{First, the U test value and P-value are shown in Table~\ref{BlindMi_Pvalue}. A large U test value and P-value indicates that two sets are similar, indicating statistical insignificance. Random generation is significantly different from all three other methods; Cross-domain sample selection is more similar to random perpetuation than sample transformation.}

\highlight{Second, we show the MMD value with standard error of the mean between generated samples and real-world non-members. A smaller MMD value indicates that the generated samples are close to real-world nonmembers. Clearly, nonmember generated by the sample transformation is the closest to real-world nonmembers; those generated randomly are the farthest---the MMD value is even larger than the one of members. The reason is that random generated samples follow a uniform distribution, which are far from the member and non-member boundary.}

\subsection{\textcolor{blue}{RQ4: \sysdiff with different classifiers and kernel functions}} \label{different kernal}

\highlight{In this subsection, we evaluate the performance of \sysdiff with different internal parameters, e.g., \sysdiff with different kernel functions and \sysdiff-w/o with different initial separation classifiers. The performances of different kernel functions are in Table~\ref{tab:kernel} and the ones of different classifiers in Table~\ref{tab:iniclassifier}.}

\textcolor{blue}{[Observation RQ4-1] \textit{The Gaussian kernel outperforms other kernels in most of the cases.}}

\textcolor{blue}{As shown in Table~\ref{tab:kernel}, the Gaussian kernel outperforms other kernels in all the datasets of \sysdiff-w/ and most of the datasets of \sysdiff-w/o (except for CH-MNIST and CIFAR-100); the Laplacian kernel comes next due to its similarity to the Gaussian kernel (the former adopts L1-norm and the latter L2-norm); the linear kernel, due to its simplicity, performs the worst. }


\begin{table}[!t]
  \renewcommand{\arraystretch}{1.2}
  \setlength{\tabcolsep}{2.5pt}
  \scriptsize
    \caption{\textcolor{blue}{F1-Score (\%) with standard error of mean for different kernel functions of \sysdiff}}
    \label{tab:kernel}
	\centering
	\begin{tabular}{l|cccccc}
	    \toprule 
	     & & \textbf{Gaussian (default)} & \textbf{Laplacian} & \textbf{Linear} & \textbf{Sigmoid} & \textbf{Polynomial}  \\
	    	    \midrule
	    \multirow{7}{*}{\rotatebox[origin=c]{90}{\textbf{{\sc diff}-w/}}} 
	    & Adult & \textbf{64.2$\pm$1.59} & 60.3$\pm$0.38 & 40.7$\pm$0.20& 51.1$\pm$0.41& 58.4$\pm$1.02 \\
	    & EyePACS  & \textbf{77.7$\pm$0.80} & 67.3$\pm$0.31 & 71.8$\pm$0.93& 72.8$\pm$0.87& 73.9$\pm$0.88  \\
	     & CH-MNIST & \textbf{75.1$\pm$1.49} & 73.1$\pm$0.92 & 72.4$\pm$0.53& 71.3$\pm$0.71& 72.7$\pm$1.20 \\
	     & Location & \textbf{86.2$\pm$0.90} & 85.1$\pm$2.42 & 83.4$\pm$0.98 & 79.8$\pm$1.52 & 76.7$\pm$0.17\\
	     & Purchase-50 & \textbf{78.0$\pm$0.31} & 68.9$\pm$0.50 & 75.8$\pm$0.61& 71.1$\pm$1.05& 66.0$\pm$0.99\\
	     & Texas & \textbf{85.5$\pm$0.80} & 83.6$\pm$0.47 & 81.2$\pm$0.29& 80.9$\pm$0.49& 81.9$\pm$1.72 \\
	     & CIFAR-100 & \textbf{93.9$\pm$0.63} & 93.3$\pm$0.79 & 87.9$\pm$1.09& 86.9$\pm$1.02& 90.1$\pm$0.83\\
	     & Birds-200 & \textbf{96.8$\pm$0.09} & 91.9$\pm$1.32 & 95.7$\pm$1.06& 94.4$\pm$1.31& 93.9$\pm$0.96\\
	     \midrule
	    \multirow{7}{*}{\rotatebox[origin=c]{90}{\textbf{{\sc diff}-w/o}}} 
	    & Adult  & \textbf{62.7$\pm$1.12} & 52.2$\pm$0.74 & 50.1$\pm$0.32& 48.9$\pm$0.63& 57.1$\pm$1.83\\
	    & EyePACS  & \textbf{75.0$\pm$1.40} & 72.9$\pm$0.65 & 69.4$\pm$0.19& 69.2$\pm$0.28& 70.1$\pm$0.53 \\
	     & CH-MNIST & 75.1$\pm$1.89 & \textbf{75.7$\pm$2.22} & 72.9$\pm$1.23& 71.9$\pm$0.84& 73.0$\pm$1.82\\
	     & Location & \textbf{83.3$\pm$0.57} & 81.2$\pm$1.89 & 76.4$\pm$0.67& 77.4$\pm$2.15 & 72.1$\pm$0.08\\
	     & Purchase-50 & \textbf{76.5$\pm$0.25} & 66.1$\pm$0.67 & 74.9$\pm$0.09& 74.5$\pm$0.38& 76.5$\pm$1.12\\
	     & Texas & \textbf{80.7$\pm$2.37} & 76.2$\pm$1.24 & 74.1$\pm$0.80& 74.7$\pm$0.79& 75.8$\pm$1.02\\
	     & CIFAR-100 & 92.1$\pm$1.15 & \textbf{92.8$\pm$1.32} & 82.9$\pm$0.33& 80.9$\pm$0.36& 88.9$\pm$0.86 \\
	     & Birds-200 & \textbf{96.2$\pm$0.26} & 96.0$\pm$0.34 & 95.7$\pm$0.83& 94.1$\pm$0.51& 94.4$\pm$1.02\\
	    \bottomrule
	 \end{tabular}
\end{table}

\textcolor{blue}{[Observation RQ4-2] \textit{The threshold classifier outperforms other initial sample separation classifiers for \sysdiff-w/o.}}

\textcolor{blue}{We evaluate three initial sample separation classifiers. The threshold classifier (``Threshold'') is a separation based on the highest probability score, among which we select the 1,000 lowest ones as our nonmembers. The others are two different clustering algorithm including K-means and Agglomerative Clustering.}

\highlight{Table~\ref{tab:iniclassifier} shows that the ``Threshold'' is the worst for initial F1-score but the best after \sysdiff-w/o. The reason is the ``Threshold'' only selects a few samples with a high probability to be nonmembers.  Since ``Threshold'' left out many nonmembers, the initial F1-score is relatively low; at the same time, a high quality nonmember set also helps \sysdiff-w/o to achieve a relatively good performance. The results of K-means and Agglomerative Clustering are similar. The initial F1-scores are higher than ``Threshold''; however, since there does not exist a set with high quality nonmembers or  members, the performance of \sysdiff-w/ is relatively lower.}

\begin{figure}[!t]
\centering
\includegraphics[width=0.95\linewidth]{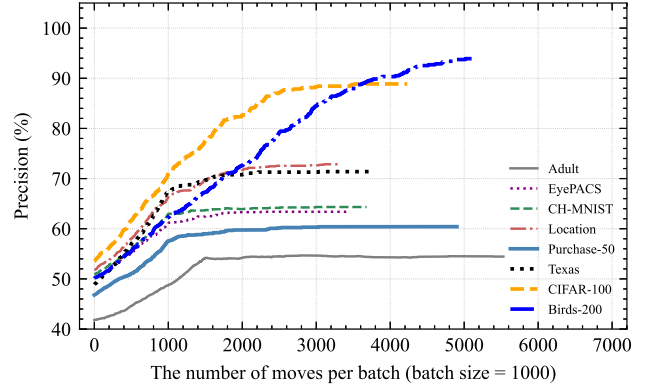}
\graphicspath{{../eps/}}
\DeclareGraphicsExtensions{.eps}\vspace{-0.15in} 
\caption{\textcolor{blue}{Precision vs. \# of moves per batch for \sysdiff-w/o.}
 } 
\label{fig:pre_diff}
\end{figure}

\begin{figure}[!t]
\centering
\includegraphics[width=0.95\linewidth]{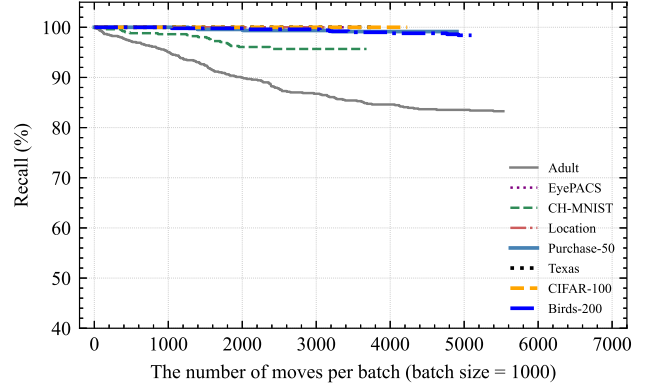}
\graphicspath{{../eps/}}
\DeclareGraphicsExtensions{.eps}\vspace{-0.15in}
\caption{\textcolor{blue}{Recall vs. \# of moves per batch for \sysdiff-w/o.}} 
\label{fig:recall_diff}
\end{figure}

\begin{figure}[!t]
\centering
\includegraphics[width=0.95\linewidth]{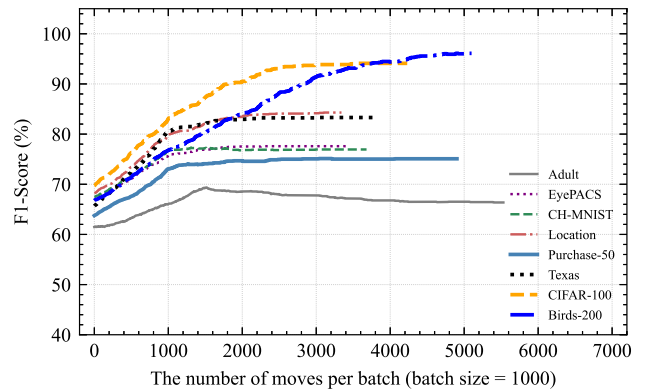}
\graphicspath{{../eps/}}
\DeclareGraphicsExtensions{.eps}\vspace{-0.15in}
\caption{\textcolor{blue}{F1-Score vs. \# of moves per batch for \sysdiff-w/o.}}
\label{fig:f1_diff}
\end{figure}

\highlight{We also show the precision, recall and F1-score of \sysdiff-w/ (with ``Threshold'' as the classifier) as the number of iterations increases in Figures~\ref{fig:pre_diff}, \ref{fig:recall_diff}, and~\ref{fig:f1_diff}. The recall starts from a point that is very close to 1 and drops as the number of iterations; by contrast, the precision increases steadily together with the F1-score. It is worth noting that the recalls of Adult and CH-MNIST drop the most compared with other datasets because members are more similar to nonmembers in target models trained from these two datasets.}


\begin{table}[!t]
  \renewcommand{\arraystretch}{1.2}
  \setlength{\tabcolsep}{2.5pt}
  \scriptsize
    \caption{\textcolor{blue}{F1-Score (\%) with standard error of mean for different rough sample separation classifiers for \sysdiff-w/o.}}
    \label{tab:iniclassifier}
	\centering
	\begin{tabular}{c|c|c|c|c|c|c}
	    \toprule 
	    \multirow{2}{*}{Dataset}& \multicolumn{2}{c|}{\textbf{Threshold (default)}} & \multicolumn{2}{c|}{\textbf{K-means}} & \multicolumn{2}{c}{\textbf{Agg. Clustering}} \cr
	    \cline{2-7}
	    & initial & + diff-w/o. & initial & + diff-w/o. & initial & + diff-w/o.\cr
	    \midrule
	    Adult & 60.2$\pm$0.04 & \textbf{62.7$\pm$1.12}& 55.1$\pm$1.75& 60.1$\pm$1.02 & 58.7$\pm$0.90& 59.4$\pm$0.23\cr
	    EyePACS & 70.6$\pm$0.58 & \textbf{75.0$\pm$1.40} & 70.0$\pm$1.15& 74.9$\pm$0.23 & 70.0$\pm$1.15& 73.0$\pm$0.50\cr
	    CH-MNIST & 73.2$\pm$0.71 & \textbf{75.1$\pm$1.89} & 70.3$\pm$0.18& 72.0$\pm$2.46 & 69.8$\pm$0.21& 76.3$\pm$1.41\cr
	    Location & 76.9$\pm$0.00 & \textbf{83.3$\pm$0.57}& 74.2$\pm$0.43& 82.2$\pm$4.84& 70.6$\pm$0.86& 81.3$\pm$0.06\cr
	    Purchase-50 & 69.0$\pm$0.00 & \textbf{76.2$\pm$0.25}& 73.6$\pm$0.28& 74.2$\pm$1.23 & 72.7$\pm$0.90& 73.3$\pm$0.66\cr
	    Taxes & 68.9$\pm$0.03 & \textbf{80.7$\pm$2.37} & 71.4$\pm$0.33 &77.0$\pm$1.51& 70.6$\pm$0.49& 79.4$\pm$1.47\cr
	    CIFAR-100 & 68.8$\pm$0.13 & \textbf{92.1$\pm$1.15} & 82.9$\pm$1.01& 87.7$\pm$0.98 & 81.1$\pm$3.20& 86.2$\pm$4.20\cr
	    Birds-200 & 71.4$\pm$0.03 & \textbf{96.2$\pm$0.26} & 92.9$\pm$0.77 &93.5$\pm$0.23& 94.7$\pm$0.99 & 96.1$\pm$0.37\cr
	    \bottomrule
	 \end{tabular}
\end{table}

%

\subsection{\textcolor{blue}{RQ5: Number of Moves, Iterations, and Execution Time of \sysdiff}}


\highlight{In this research question, we measure the time and the numbers of moves and iterations of \sysdiff to finish the inference of the target dataset.  Note that moves are atomic steps in which \sys moves a sample from $S^{prob,k}_{target}$ to $S^{prob,k}_{nonmem}$.  Then, iterations are
when \sys updates the distance between $S^{prob,k}_{target}$ and $S^{prob,k}_{nonmem}$.  The evaluation results are shown in Table~\ref{tab:time}.}

%

\textcolor{blue}{[Observation RQ5-1] \textit{The execution time and the number of moves and iterations depend on the size of the target dataset.}}

\highlight{Our first observation is that the execution time and number of moves and iterations depend on the size of the target dataset. The larger the target dataset is, the longer time and more moves and iterations it takes for \sysdiff to finish the inference. The reason is that the larger size of the datasets increases the number of moves per interation, and thus increases the potential time and numbers of iterations taken by \sysdiff. }



\textcolor{blue}{[Observation RQ5-2] \textit{\sysdiff-w/o takes significantly longer time, and more moves, than \sysdiff-w/.}}

\highlight{Our second observation is that \sysdiff-w/o is generally slower than \sysdiff-w. The reason is that \sysdiff-w/o adopts bi-directional differential comparison: The moves are bi-directional and thus the number of \sysdiff-w/o is larger than  \sysdiff-w/.}

\begin{table}[!t]
  \renewcommand{\arraystretch}{1.2}
  \setlength{\tabcolsep}{2.5pt}
  \scriptsize
    \caption{\textcolor{blue}{Execution Time (second) and \# of Moves and \# of Iterations with Standard Error of mean for \sysdiff} 
    }
    \vspace{-0.05in}
    \label{tab:time}
	\centering
	\begin{tabular}{c|c|c|c|c|c|c}
	    \toprule 
        \multirow{2}{*}{Dataset}& \multicolumn{3}{c|}{\textbf{BlindMI-diff-w/.}} & \multicolumn{3}{c}{\textbf{BlindMI-diff-w/o.}} \cr
        \cline{2-7}
        & Time (s)& Moves (\#) & Iter. (\#) & Time (s) & Moves (\#) & Iter. (\#) \cr
        \midrule
        Adult & 494$\pm$23 & 63,124$\pm$616 & 7,012$\pm$98 &2,530$\pm$28 & 202,407$\pm$694 & 405$\pm$14\cr
        EyePACS &224$\pm$16 & 44,838$\pm$858 & 2,818$\pm$14 &751$\pm$31 & 94,247$\pm$448 & 120$\pm$9 \cr
        CH-MNIST & 73$\pm$11& 12,181$\pm$386 & 983$\pm$29 &293$\pm$28 & 25,061$\pm$429 & 30$\pm$2 \cr
        Location & 70$\pm$2 & 9,839$\pm$120 &  857$\pm$4 &271$\pm$18 & 20,659$\pm$464 & 31$\pm$2\cr
        Purchase-50 & 370$\pm$6&  48,943$\pm$373 &  4,336$\pm$114 &1,215$\pm$45 & 97,243$\pm$761 & 127$\pm$3\cr
        Texas & 313$\pm$5 &47,428$\pm$903  & 3,086$\pm$65 &781$\pm$4 & 67,379$\pm$746 & 110$\pm$7\cr
        CIFAR-100 & 238$\pm$15 & 41,128$\pm$358 &  3,051$\pm$101 &984$\pm$59 & 104,006$\pm$310 & 168$\pm$9\cr
        Birds-200 & 183$\pm$18 & 30,261$\pm$647 &  2,067$\pm$25 &842$\pm$68 & 70,109$\pm$325 & 107$\pm$2\cr
	    \bottomrule
	 \end{tabular} 
\end{table}

\textcolor{blue}{[Observation RQ5-3] \textit{The total number of iterations depends on the batch size.}}

\highlight{Our third observation is that the batch size determines the number of iterations: That is why \sysdiff-w/ with a batch size as 20 takes more iterations than \sysdiff-w/o with a batch size as 1,000. Specifically, when the batch size is small, the number of batches is large, but the number of iterations per batch does not differ much, leading to a large number of iterations in total.  }

\textcolor{blue}{[Observation RQ5-4] \textit{The distance between two sets increases as the number of moves per batch.}}

\begin{figure}[!t]
\centering
\includegraphics[width=0.95\linewidth]{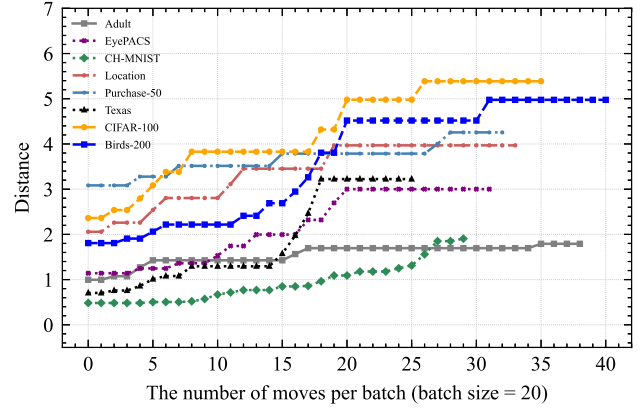}
\graphicspath{{../eps/}}
\DeclareGraphicsExtensions{.eps} \vspace{-0.15in}
\caption{\textcolor{blue}{Distance vs. \# of iterations per batch for \sysdiff-w/.}} 
\label{fig_distance_w/} 
\end{figure}

\begin{figure}[!t]
\centering
\includegraphics[width=0.95\linewidth]{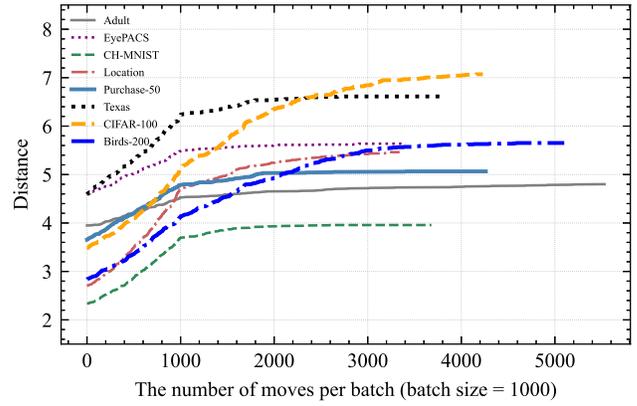}
\graphicspath{{../eps/}}
\DeclareGraphicsExtensions{.eps}\vspace{-0.15in}
\caption{\textcolor{blue}{Distance vs. \# of iterations per batch for \sysdiff-w/o.}} 
\label{fig_distance_w/o}
\end{figure}

\highlight{Our last observation is on the distance between two sets vs. the number of moves per batch as shown in Figures~\ref{fig_distance_w/}  (\sysdiff-w/) and~\ref{fig_distance_w/o} (\sysdiff-w/o). 
 Note that only a move that increases the distance is a valid one between two sets; otherwise, the sample is kept in the original set.}



\subsection{RQ6: \sys with Different Configurations}

In this subsection, we evaluate \sys with different configurations, including different nonmember-to-member ratios (Bargav et al.~\cite{realisticMI}) and different prediction classes. The evaluations are performed under the blackbox setting as many attacks require ground-truth labels. 

\subsubsection{Different Nonmember-to-member Ratios}

\begin{figure}[!t]
\centering
\includegraphics[width=0.95\linewidth]{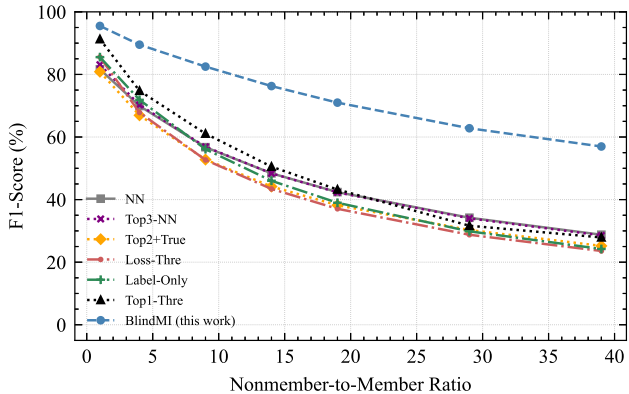}
\graphicspath{{../eps/}}
\DeclareGraphicsExtensions{.eps} \vspace{-0.15in}
\caption{F1-Score of Various Attacks vs. Nonmember-to-Member Ratio on CIFAR-100.}
\label{fig_cifer100_ratio} 
\end{figure}

In this part, we evaluate the F1-score of \sys and existing attacks when the nonmember-to-member in the target dataset changes. Specifically, we follow Bargav et al.~\cite{realisticMI} to adjust the Nonmember-to-Member ratio $r$ and measure the F1-score.  The underlying rational behind the introduction of $r$ is that a practical target dataset usually has a small number of members and a large number of nonmembers. Our evaluation results based on CIFAR-100 are shown in Figure~\ref{fig_cifer100_ratio}.

[Observation RQ6-1] \textit{While the performance of all MI attacks degrades as the nonmember-to-member ratio ($r$) increases, \sys is the slowest among all and significantly outperforms existing attacks at a large $r$ value. }

This observation shows the practicability of \sys under real-world settings. All other attacks in the literature drops logarithmically as $r$ increases, while the performance decrease of \sys is stable.  That is, the performance of existing attacks drops below 50\% when $r$ is larger than 10, while the performance \sys is still above 50\%, i.e., 57.5\% (35\% than the state-of-the-art), when $r$ equals to 39. 

\subsubsection{Different Prediction Classes} \label{vs.class}
\begin{figure}[!t]
\centering
\includegraphics[width=0.95\linewidth]{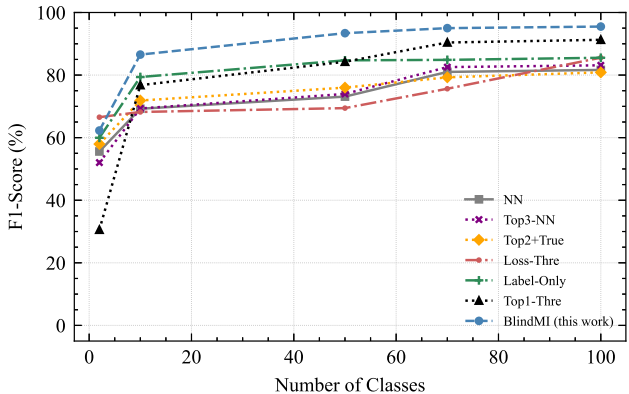}
\graphicspath{{../eps/}}
\DeclareGraphicsExtensions{.eps}\vspace{-0.15in}
\caption{F1-Score of Various Attacks vs. \# of classes on CIFAR.} \vspace{-0.15in}
\label{fig_Classes}
\end{figure}

In this part, we evaluate the F1-score of \sys and all other attacks when the number of classes in the target model increases. The experiment settings are as follows. We divide the entire CIFAR-100 datasets into subsets with 2, 10, 50, 70, and 100 classes and then launch MI attacks against target models trained from these subsets.  The F1-scores of these attacks are shown in Figure~\ref{fig_Classes} and our observation is as follows. 



\textcolor{blue}{[Observation RQ6-2] \textit{The performance of all MI attacks, including \sys, increase as the number of classes in the target model and this performance boost is more significant when the number of classes is small.}}

\textcolor{blue}{Figure~\ref{fig_Classes} shows a steady improvement of all MI attacks as the number of classes.  The reason is that when the dataset has more classes, the target model tends to generalize less, thus being more vulnerable to MI attacks. This can also explain why target models trained on CIFAR-100 and Birds-200 are more vulnerable compared to other datasets. } 

\textcolor{blue}{It is also worth noting that Top1-threshold performs the worst among all the MI attacks when the number of classes equals two, but the performance improves when the number becomes 100.  That is, the top one probability score contains more information as the number of classes increases.  We believe that this may be due to the fact that  when the total probability is shared by more classes, one can infer more membership information from the top probability. }


\section{A Discussion on Potential Defenses}
\label{section_VI}

In this section, we discuss potential defenses.  There are two possible venues of defenses as we have seen in the literature. 

\begin{icompact}
\item {Limiting adversary's access to the target model.} \hspace{0.05in} The first method is to limit the adversary's access to the target model: (i) restricting the number of probes and also which samples can be probed, and (ii) providing only the predicted class information as an output.  The former will restrict \sys to \sysdiff-w/o, which performs a little bit worse than \sysdiff-w/. The latter will reduce \sys to the Label-only attack in the blackbox setting. 
\item {Improving the robustness of the target model.} \hspace{0.05in} The second method is to improve the model robustness to MI attacks via different methods evaluated in Section~\ref{RQ2}. The differential privacy-based approach is likely the best method for defending against \sys in the literature.  A combination of existing attacks may also be possible and this is left as a future work for our study. 
\end{icompact}

%
%
%

\section{Related Work} \label{Related Work}


Machine learning  is vulnerable to different privacy attacks including model inversion~\cite{fredrikson2014privacy,modelInversion}, membership inference~\cite{earliest}, property inference~\cite{Ateniese15,Ganju18}, as well as model and hyperparameter stealing~\cite{tramer2016stealing,stealingPara}. Our work studies membership inference (MI) attacks. We describe related work on MI attacks and  defenses in Section~\ref{subsec:relatedattack} and~\ref{subsec:relateddefense}. 




\subsection{Existing Membership Inference (MI) Attack\highlight{s}} \label{subsec:relatedattack}


Membership inference attacks originate back to 2008, when, Homer et al.~\cite{first} first proposed a MI attack on biological data, whereby an adversary could infer whether a data sample belonged to a genome-based study knowing only parts 
 the genome and summary statistics. Then, in 2017, Shokri et al.~\cite{earliest} \highlight{proposed} the first modern MI attack against deep neural networks with a shadow model and a binary attack classifier. 
 
 Prior attack methods include the following. Salem et al.~\cite{NDSS} proposed several MI attacks. For example, the Top3-NN attack of Salem et al., a variant NN Attack, picks the top three largest values from all confidence scores to train an MI classifier.  For another example,  the Top1-Threshold attack of Salem et al. compares the top feature from the output probability distribution with a threshold and classified the sample as member if the top feature is larger than a threshold. Similarly, Yeom et al.~\cite{globalloss} also proposed two attacks with the help of ground-truth labels: the first label-only attack compares the ground truth label with predicted, and the second loss-threshold attack computes cross-entropy loss and compares the computed loss \highlight{with} the average loss of all training samples. 
  As a comparison, \sys is an attack that does not need a shadow model but also extracts complex membership semantics via probing only.  Our evaluation shows that \sys outperforms existing attacks under different adversarial settings.

Researchers have also proposed theories on MI attacks.  Sablayrolles et al.~\cite{BOS_attack} proposed an optimal strategy for MI attacks using a probabilistic framework that consists of both Bayesian learning and noisy training.  They showed that optimal attacks only depend on the loss function, and thus blackbox attacks could be as good as whitebox attack. \sys actually proves the effectiveness of blackbox attacks. 

 In addition to attacks on classification models, researchers also have proposed MI attacks~\cite{logan} on generative models~\cite{GAN} and those~\cite{white-box} on federated learning.  As a comparison, \sys is an attack on single classification models rather than generative models or federated learning.

\subsection{Existing Defense\highlight{s}} \label{subsec:relateddefense}

We now describe existing defenses of MI attacks\cite{previous_work1,previous_work2,minmax,previous_work4}, especially those on classification models.  
  Note that while existing defenses can prevent some existing MI attacks with a reasonable performance, our evaluation shows that \sys can still infer membership with a reasonable F1-score, e.g., over 60\%.

%
%
\subsubsection{Regularization}

Researchers have proposed to improve privacy against MI attacks via different types of regularization. 
%
 For example, Salem et al.~\cite{globalloss} demonstrated two effective method of defending MI attacks, namely dropout and model stacking.       The former randomly deletes a fixed proportion of edges in a fully connected neural network model  to improve model robustness; the latter constructs a target model with multiple different machine learning models stacked together.      For another example, Shokri et al.~\cite{earliest} 
 adopted $L_2$-norm standard regularization with a polynomial in the model's loss function to penalize large parameters.  
  Nasr et al.~\cite{minmax} introduced a min-max game mechanism to train models with membership privacy, which ensures indistinguishability between the predictions of a model on its training data and other data points from the same distribution. 
 This strategy 
 acts as an adversarial regularizer that generalizes the model. 
 In addition, Li et al.~\cite{mixup} proposed to close the generalization gap by matching the training and validation accuracies.  Specifically, they adopted a new set regularizer, called the Maximum Mean Discrepancy, between the softmax output empirical distributions of the training and validation sets during training. 

\subsubsection{Adversarial Example}

Another direction is to borrow ideas from adversarial machine learning and generate an adversarial example for the  inference model controlled by the adversary.  For example, Jia et al.~\cite{memguard} introduced a new defense, called MemGuard, by adding noise to confidence score output from target models, thus fooling a binary classifier.  Unlike previous adversarial examples~\cite{adver1,adver2,adver3,adver4,adver5,adver6,adver7,adver8}, MemGuard calculates the gradient of the loss function to find an appropriate noise and guarantee the utility loss to be zero.


\subsubsection{Privacy Enhancement}

  Many differential privacy based defenses~\cite{first_dp,dp_1,dp_2} add noise to the objective function that is used to learn a model or the gradients during optimizing the objective function. Shokri et al.~\cite{dp_paper} designed a differential privacy method for collaborative learning of DNNs. Cao et al.~\cite{unlearning:sp15} showed that privacy-related data samples can be unlearned to improve model privacy. 
  


\section{Conclusion} \label{Conclusion}


In this paper, we present a novel MI attack, called \sys, which adopts differential comparison moving samples in between two sets and making inference decisions. One of the key insights used here is that, moving a member from a mostly member dataset to a mostly non-member one will decrease the distance in feature space between two sets and vice-versa.  We implement three versions of \sys, \syssvm (relying on one-class SVM), \sysdiff-w/ (relying on generation of nonmembers), and \sysdiff-w/o (relying on rough separations of members and non-members). Our evaluation shows that \sys outperforms existing state of the art attacks, against not only a variety of DNN architectures, but also against DNNs with state of the art defenses deployed. 




%
%



%

\section*{Acknowledgment}

We want to thank anonymous
reviewers for their helpful comments and feedback. This work
was supported in part by the Johns Hopkins
University Institute for Assured Autonomy with grant IAA 80052273,
 National Science Foundation (NSF)
grant CNS-18-54000 and CNS-19-37786, as well as an IBM Faculty Award. The views and conclusions contained
herein are those of the authors and should not be interpreted as
necessarily representing the official policies or endorsements,
either expressed or implied, of NSF.

\bibliographystyle{IEEEtranS}
\bibliography{paper-main}

\end{document}